\documentclass[apjl]{emulateapj}

\usepackage{graphicx}

\shorttitle{Dark matter burning in star clusters}
\shortauthors{J. Casanellas and I. Lopes}

\begin{document}

\slugcomment{To be published in The Astrophysical Journal Letters}

\title{Signatures of dark matter burning in nuclear star clusters}

\author{Jordi Casanellas\altaffilmark{1,3}, Il\'\i dio Lopes\altaffilmark{1,2,4}}

\altaffiltext{1}{Centro Multidisciplinar de Astrof\'{\i}sica, Instituto Superior T\'ecnico, Av. Rovisco Pais, 1049-001 Lisboa, Portugal} \altaffiltext{2}{Departamento de F\'\i sica, Universidade de \'Evora, Col\'egio Luis Ant\'onio Verney, 7002-554 \'Evora - Portugal} 
\altaffiltext{3}{E-mail: jordicasanellas@ist.utl.pt} \altaffiltext{4}{E-mail: ilidio.lopes@ist.utl.pt}  

\begin{abstract}
In order to characterize how dark matter (DM) annihilation inside stars changes the aspect of a stellar cluster we computed the evolution until the ignition of the He burning of stars from 0.7 M$_{\odot}$ to 3.5 M$_{\odot}$ within halos of DM with different characteristics. We found that, when a cluster is surrounded by a dense DM halo, the positions of the cluster' stars in the H-R diagram have a brighter and hotter turn-off point than in the classical scenario without DM, therefore giving the cluster a younger appearance. The high DM densities required to produce these effects are expected only in very specific locations, such as near the center of our Galaxy. In particular, if DM is formed by the 8 GeV WIMPs recently invoked to reconcile the results from direct detection experiments, then this signature is predicted for halos of DM with a density $\rho_{\chi}=3\cdot10^5\;$GeV cm$^{-3}$. A DM density gradient inside the stellar cluster would result in a broader main sequence, turn-off and red giant branch regions. Moreover, we found that for very high DM halo densities the bottom of the isochrones in the H-R diagram rises to higher luminosities, leading to a characteristic signature on the stellar cluster. We argue that this signature could be used to indirectly probe the presence of DM particles in the location of a cluster.
\end{abstract}

\keywords{dark matter - Galaxies: star clusters - Galaxy: center - Hertzsprung-Russell diagrams - Stars: fundamental parameters}

\section{Introduction}
An unambiguous discovery of the particle nature of dark matter (DM) would have to come simultaneously from a variety of experiments and observations \citep{art-Bertone2010Nature}. Positive results from direct detection experiments \citep{chap-CerdenoGreen2010,art-Patoetal2010arXiv} and the hypothetical evidence of the existence of new particles from colliders \citep{art-Bertone2010PhRevD} must be complemented by indirect methods, such as the detection of DM annihilation products \citep{art-Trottaetal2009JCAP,art-Scottetal2010JCAP,art-BernalPalomares2010} or the observation of a peculiar signature in the solar neutrinos attributed to the effect of captured DM particles \citep{art-Taosoetal2010PhRvD,art-LopesSilk2010Science}.

In the last years many works studied the effects of WIMP DM on stellar evolution \citep{let-Spolyaretal2008,art-BertoneFairbairn2008,art-Iocco2008,art-YoonIoccoAkiyama2008,art-Taosoetal2008PhRevD,art-RipamontiIocooetal2010MNRAS,art-GondoloHuhKimScopel2010JCAP,art-SivertssonGondolo2010,art-LavallazFairbairn2010PhRvD,art-ZackScottIoccoetal2010ApJ,art-KouvarisTinyakov2010,art-Yuanetal2011} as a promising complementary way to investigate the nature of DM. Remarkably, it has also been argued that the seismological analysis of the stellar oscillations could be used to detect the signature of captured DM particles in the Sun \citep{art-Cumberbatchetal2010PhRevD,art-LopesSilk2010ApJL} and in other sun-like stars in environments with very high DM densities \citep{art-CasanellasLopes2010MNRAS}. All these studies require DM particles to interact with a non-zero nuclear scattering cross section.

In this work we are interested in the global behavior of a large group of stars instead of being concerned with the influence of DM on a single star, whose observation would require a higher precision. We address the question of how a dense halo of DM particles changes the properties of an embedded cluster of stars. As we will show, the annihilation of captured DM particles inside the stars leaves strong signatures in the stellar cluster when compared with a classical cluster without DM. The high DM densities required to produce measurable effects on the cluster restrict our study to the nuclear star clusters, present in the centers of galaxies, where the highest DM densities are expected. Our description of the cluster isochrones provides an indirect way to probe the presence of DM particles in the location of the cluster, as the signatures we describe here are difficult to attribute to other processes.

This letter is organized as follows: the physics beyond the stellar models and the capture and annihilation of DM particles is briefly described in Section \ref{sec-StellDMphysics}; the effects of DM on stellar evolution are characterized in Section \ref{sec-stellevolution}; in Section \ref{sec-DMcluster} the properties of a cluster embedded in a dense DM halo are compared with those of a classical cluster; finally, we conclude in Section \ref{sec-discussion} with a brief discussion of our results.

\section{Stellar and dark matter physics}
\label{sec-StellDMphysics}
To compute our stellar models we used the stellar evolution code CESAM \citep{art-Morel1997}. This code has an up-to-date and very refined microscopic physics, tested against helioseismic data \citep{art-TurckLopes1993ApJ,art-Turck-Chiezeetal2010ApJ}. Our stellar models were evolved from the zero age main sequence (ZAMS) (although some of them were also evolved from the pre-main sequence phase to check that both approaches led to similar results), at constant mass, with a metallicity $Z=0.019$ and an initial helium mass fraction $Y=0.273$ similar to the solar ones. The initial abundance of the other elements was set equal to the solar composition. The mixing-length parameter was set by calibrating a solar model with an accuracy of $10^{-5}$ on the solar radius and luminosity. The performance of our code in the range of masses (0.7 M$_{\odot}$ to 3.5 M$_{\odot}$) and evolutionary stages studied in this work was successfully tested by comparing our computed isochrones with those of \cite{art-Girardietal2000A&AS}.

The stars computed in this work are embedded in a dense halo of DM. To account for the impact of the DM particles on the stars we considered that some of the DM particles that populate the halo are gravitationally captured by the stars and accumulate in their interior. The number of captured DM particles was computed using the integral expression of \cite{art-Gould1987}, as implemented in \cite{art-GondoloEdsjoDarkSusy2004}. Note that, for the capture process to be efficient, the DM particles are assumed to have a non-negligible scattering cross section with baryons $\sigma_{\chi}$, which we chose to be smaller than the present limits from direct detection experiments: $\sigma_{\chi,SI}=10^{-44}\;$cm$^2$ \citep{art-CDMS2010Sci} and $\sigma_{\chi,SD}=10^{-38}\;$cm$^2$ \citep{art-COUPP2011PhRvL} for a WIMP with a mass of 100 GeV. For these values of $\sigma_{\chi}$, the spin-dependent (SD) interactions with hydrogen atoms always dominate over the spin-independent (SI) ones with other stellar isotopes. 

In the capture rate ($C_{\chi}$) calculation we assumed a stellar velocity $v_{\star}=220\;$km s$^{-1}$ and a Maxwellian DM velocity distribution with a dispersion $\bar{v_{\chi}}=270\;$km s$^{-1}$. These values apply for the solar case, but are certainly inaccurate for a nuclear cluster. For instance, stars with velocities as high as 400 km s$^{-1}$ are observed near the Galactic center (GC) \citep{art-LuGhezetal2009}. In this case the capture rate would be reduced by a factor of 6 (for a more thorough analysis of how $C_{\chi}$ varies for different stellar and DM characteristics see \cite{art-LopCasEug2011PhRevD}). At the same time, it is complex to model the DM velocity distribution in the GC, as the motion of the DM particles is strongly influenced by the gravitational potential of the stars and the central black hole. Interestingly, \cite{art-Scottetal2009MNRAS} tested other DM velocity distributions with the aim of grasping the possible variations on $C_{\chi}$. When a non-Gaussian distribution (designed to fit a N-body simulation of a Milky Way-size DM halo) was implemented, the capture rate was boosted by a factor of 3-5. On the other hand, the same authors found that the truncation of the isothermal distribution at the local escape velocity reduces $C_{\chi}$ by a factor of 2. The same order of uncertainty on $C_{\chi}$ is expected in the cases presented in the present work.

After some scatterings, the DM particles sink to the core of the star and rapidly thermalize with stellar matter. The number of DM particles in the stellar core increases until their self-annihilation rate balances the capture rate. This equilibrium is reached in a time scale below 10$^4$ yr for all cases studied here. Thus, the annihilation of DM particles provides a new source of energy which contributes to the total luminosity of the star according to \citep{art-SalatiSilk1989}:
\begin{equation}
L_{\chi} = f_\chi  \;m_{\chi}\;C_{\chi}\;,
\label{eq_lum}
\end{equation}
where $m_{\chi}$ is the mass of the DM particles, and $f_\chi=2/3$ to take into account that one third of the energy may escape the star in the form of neutrinos \citep{art-Ioccoetal2008}. This energy is injected to the stellar models following the thermal distribution of the DM particles, which characteristic radius is below 2\% and 7\% of the stellar radius for $m_{\chi}=100\;$GeV and 8 GeV respectively. The total input of energy from DM annihilation, and thus also its impact on stellar evolution, will depend mainly on the product $\rho_{\chi}\sigma_{\chi}$.

\section{Stellar evolution within dense DM halos}
\label{sec-stellevolution}
The hydrostatic equilibrium (the balance between pressure and gravity) achieved by a star within a dense DM halo differs from the one reached in the classical picture due to the new source of energy added to the classical thermonuclear energy sources. This fact leads to three main consequences that will influence the characteristics of the whole cluster:

1. \textit{Slowing of the evolutionary speed}: The central temperature of stars that evolve within dense DM halos is lower than that of classical stars due to their negative heat capacity. Another simple way to understand this is to imagine a forming star in the pre-main sequence. The cloud of gas that forms the proto-star shrinks, increasing its central temperature until the gravitational collapse is balanced by the thermonuclear reactions; if another source of energy helps to compensate gravity, the hydrostatic equilibrium is reached earlier, when the central temperature is lower. Therefore, stars within dense DM halos burn hydrogen at a lower rate, slowing down their evolution through later phases. For example, a star of 1 M$_{\odot}$ will spend more than $20$ Gyr in the main sequence (MS) if it evolves in a DM halo of density $\rho_{\chi}=2\cdot10^9\;$GeV cm$^{-3}$ (assuming $\sigma_{\chi,SD}=10^{-38}\;$cm$^2$, although other values of $\rho_{\chi}$ and $\sigma_{\chi,SD}$ can be considered, leading to the same effects as long as the product $\rho_{\chi}\sigma_{\chi}$ is kept constant). This is a significant difference from the classical picture, in which a star as the Sun is expected to exhaust its hydrogen core in less than 10 Gyr. As shown in earlier works \citep{art-SalatiSilk1989}, the more massive the star is, the less it is affected by WIMP annihilation. Considering the same DM halo of the previous example, a star of 3 M$_{\odot}$ won't be affected.

\begin{figure}[!t]
\centering
  \includegraphics[scale=1.]{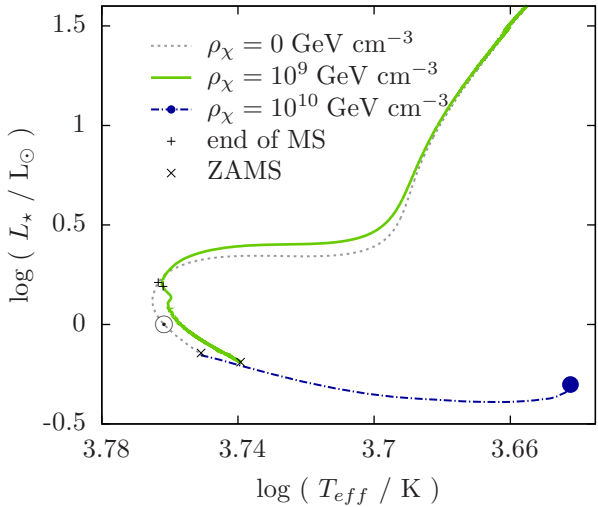}
\caption{Tracks on the H-R diagram of stars of 1 M$_{\odot}$ that evolved in halos with different DM densities. The blue point indicates a stationary state reached by a star only powered by DM burning. We considered DM particles with a mass $m_{\chi}=100\;$GeV and a spin-dependent scattering cross section with protons $\sigma_{\chi,SD}=10^{-38}\;$cm$^2$.}
\label{fig-diffHRm1}
\end{figure}

2. \textit{Different paths on the H-R diagram}: Since DM burning accounts for at least one third of the total energy, the balance will be reached with a larger radius and a lower effective temperature than in the classical picture \citep{art-Fairbairnetal2008}. Therefore, stars that evolve in dense DM halos follow slightly different paths in the H-R diagram. We found that, in addition to the different paths followed during the MS, which was already reported in previous works \citep{art-CasanellasLopes2009ApJ}, stars follow brighter tracks during the red giant branch (RGB). This feature is illustrated in Figure \ref{fig-diffHRm1}. Even if the difference in the paths is remarkable, its effect on the cluster is small compared with the slowing of the evolutionary speed.

3. \textit{Stationary states}: For extremely high DM densities, stars are powered only by the energy from DM annihilation. Whether the star was formed in this environment or arrived there a posteriori, it will reach a state of equilibrium in the Hyashi track, far from the MS where most stars are found \citep{art-CasanellasLopes2009ApJ}. In this case the star is fully convective and remains in the same position in the H-R diagram as long as there are DM particles to be captured in the halo (an illustrative example is shown in Figure \ref{fig-diffHRm1}). 

\section{Global structure of a stellar cluster within a dense DM halo}
\label{sec-DMcluster}
It is naturally expected then, that stellar clusters are affected by DM halos, since their basic constituents, namely stars, are themselves affected. The main reason is the fact that stars with lower masses evolve slower in dense DM halos. This effect is not noticeable for young clusters since in these clusters low-mass stars are still in the MS and the more massive ones, which are evolving through the RGB, are not affected by the presence of DM. However, in old clusters the RGB may be populated by stars that evolved slower, consequently making the cluster look younger than its real age. Moreover, the fact that low-mass stars within dense DM halos follow brighter paths in the RGB than classical stars contributes to amplify this effect. 

\begin{figure}[!t]
\centering
\includegraphics[scale=1.0]{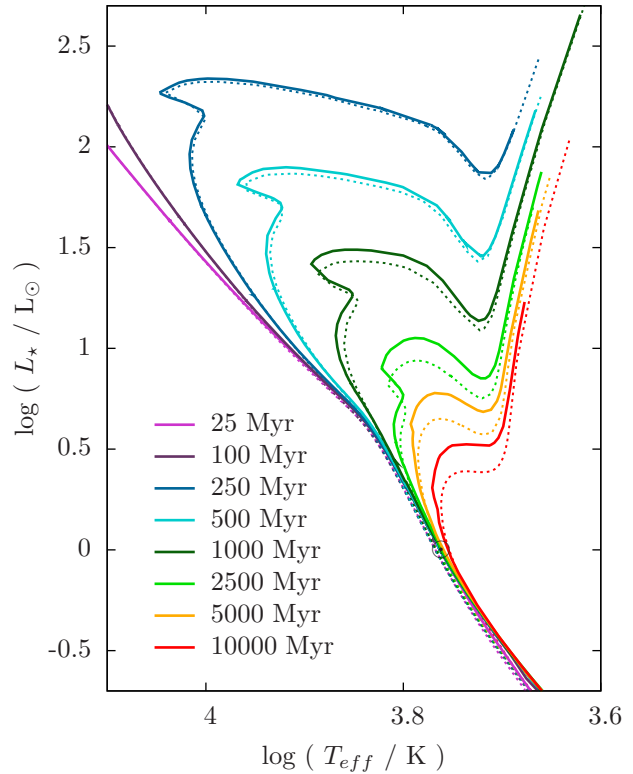}
\caption{Isochrones for a cluster of stars with masses between 0.7 M$_{\odot}$-3.5 M$_{\odot}$ that evolved in a halo of DM with a density $\rho_{\chi}=10^9\;$GeV cm$^{-3}$ (continuous lines) and for the same cluster in the classical scenario without DM (dashed lines). We considered DM particles with a mass $m_{\chi}=100\;$GeV and a spin-dependent scattering cross section with protons $\sigma_{\chi,SD}=10^{-38}\;$cm$^2$.}
\label{fig-isoe09}
\end{figure}

In order to distinctly illustrate the younger appearance of a cluster when embedded in a dense DM halo, we computed the isochrones (the track drawn by the positions in the H-R diagram of all stars with different masses at a given age) of stellar clusters in different situations. Figure \ref{fig-isoe09} shows the isochrones we obtained for a cluster evolving in a halo of DM with a density $\rho_{\chi}=10^9\;$GeV cm$^{-3}$ (continuous lines) together with those obtained without the influence of DM (dashed lines). When the isochrones of $\geq 1000\;$Myr in both situations are compared, we see that indeed the cluster within a dense DM halo looks younger, with a brighter and hotter turn-off point and a brighter RGB. In this case the turn-off and RGB are populated by more massive stars than in the classical scenario, because they took longer to burn out their hydrogen core and to leave the MS. It is almost impossible to distinguish both clusters at ages $\leq 500\;$Myr.

When even higher DM densities are considered (or, equivalently, larger WIMP-on-nucleon scattering cross sections) the characteristics of the cluster change dramatically. In addition to the previously described effect (which will now be visible for younger clusters, because at higher DM densities more massive stars will be affected), another strong signature of the presence of DM in the halo arises when looking at the position of stars with lower masses. These stars, which are mostly fueled by the energy from DM annihilation, go back in the Hyashi track and reach positions in the H-R diagram which were normally occupied only by forming stars in their way to the MS. Consequently, the bottom of the isochrones, corresponding to the lower mass stars, rises to higher luminosities, giving the cluster a very characteristic appearance. This peculiar signature is a strong indication of the presence of high concentrations of DM in a stellar cluster. 

\begin{figure}[!t]
\centering
\includegraphics[scale=1.0]{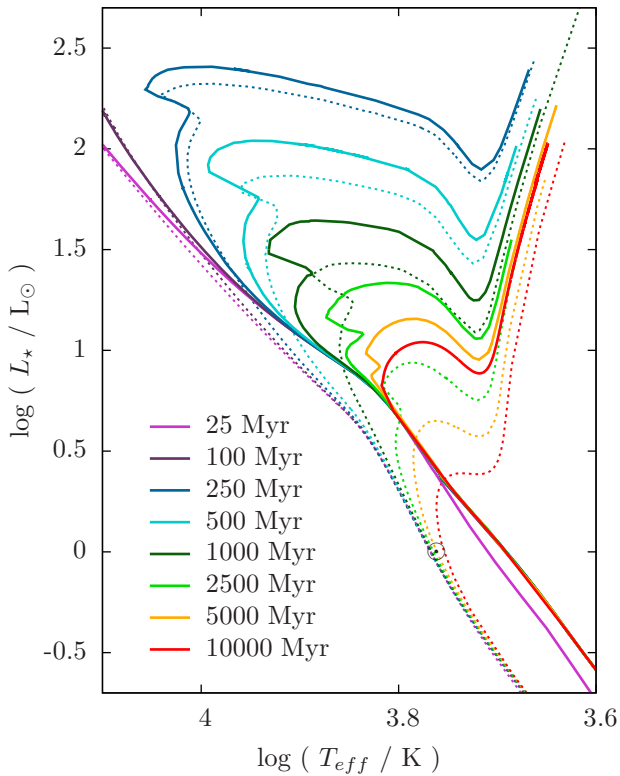}
\caption{Isochrones for a cluster of stars with masses between 0.7 M$_{\odot}$-3.5 M$_{\odot}$ that evolved in a halo of DM with a density $\rho_{\chi}=10^{10}\;$GeV cm$^{-3}$ (continuous lines) and for the same cluster in the classical scenario without DM (dashed lines). The post-MS segment of the 10 Gyr isochrone is a conservative estimation (a lower limit on luminosity) of the true isochrone. We considered DM particles with a mass $m_{\chi}=100\;$GeV and a spin-dependent scattering cross section with protons $\sigma_{\chi,SD}=10^{-38}\;$cm$^2$.}
\label{fig-isoe10}
\end{figure}

This strong signature is illustrated in Figure \ref{fig-isoe10}, where the isochrones of a stellar cluster surrounded by a halo of DM with a density $\rho_{\chi}=10^{10}\;$GeV cm$^{-3}$ are plotted. The main characteristic signature of the presence of DM is the fact that the bottom of all isochrones is more than 3 times brighter than the classical isochrones. In addition, the effect of a brighter and hotter turn-off point is now more pronounced and appreciable in clusters as young as 250 Myr. 

\begin{figure}[!t]
\centering
\includegraphics[scale=1.0]{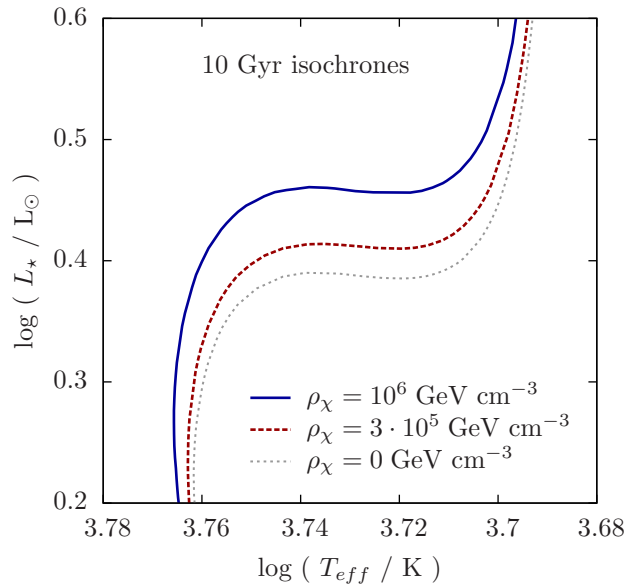}
\caption{Isochrones of 10 Gyr for clusters of stars that evolved in halos of DM with different densities. We considered DM particles with the particular characteristics that fit DAMA observations and constraints from direct detection experiments: a mass $m_{\chi}=8\;$GeV and a spin-dependent scattering cross section with protons $\sigma_{\chi,SD}=10^{-36}\;$cm$^2$.}
\label{fig-mx8}
\end{figure}

We have also considered the hypothetical scenario in which DM is formed by the low-mass WIMPs invoked to reconcile the results of DAMA with the negative results of other direct detection experiments \citep{art-Savageetal2009JCAP}. As shown in Figure \ref{fig-mx8}, if such WIMPs form most of the DM then the DM density needed to have signatures on a stellar cluster would be as low as $3\cdot10^{5}\;$GeV cm$^{-3}$. Both the low mass of these WIMPs ($m_{\chi}=8\;$GeV) and especially their large SD scattering cross section with protons ($\sigma_{\chi,SD}=10^{-36}\;$cm$^2$) contribute to producing effects on the stellar cluster at lower DM halo densities.

\section{Discussion and conclusions}
\label{sec-discussion}
We have shown that a cluster of stars that evolves in a dense halo of DM shows strong signatures in its appearance due to the self-annihilation of captured DM particles in the interior of stars. In comparison to the classical case, the cluster within a dense DM halo looks younger than its true age, due to the slower evolution of the stars when these are partially powered by DM annihilation. This is visible only for old clusters (e.g. for clusters older than 1 Gyr within a DM halo of density $\rho_{\chi}=10^9\;$GeV cm$^{-3}$), because their RGB is populated by low-mass stars, which are the type of stars most affected by DM.

Our work focuses on environments with very high DM densities, which may be present only in specific locations, such as near the centers of galaxies \citep{art-GondoloSilk1999}. In particular, considering an adiabatically contracted DM profile \citep{art-BertoneMerritt2005}, the DM densities discussed here may be found at the following distances from the GC: $\rho_{\chi}=3\times10^5\;$GeV cm$^{-3}$ at $r_{GC}\approx 1\;$pc and $\rho_{\chi}=10^{10}\;$GeV cm$^{-3}$ at $r_{GC}\approx 0.01\;$pc. The shape of the central profiles of galactic DM halos is still a topic of discussion \citep{art-deBlok2010}: while simulations predict the existence of cusps, observations favor constant-density DM cores. 

Our results indicate that the age of a cluster may be underestimated if embedded in a dense DM halo, which goes towards solving the ``paradox of youth" in the center of the Milky Way, a possibility that was first suggested by \cite{art-MoskalenkoWai2007} in the context of compact stars. However, there are many astrophysical uncertainties, such as the velocities of stars and DM particles, that may change the rate at which stars capture DM particles and therefore change the overall influence of DM on a cluster. Although our results do not explain the depletion of giants observed in the nuclear central cluster of the Milky Way \citep{art-Doetal2009ApJ,art-Buchholzetal2009A&A,art-Bartkoetal2010ApJ} they show that the influence of DM on stellar evolution must be taken into account when studying nuclear clusters.

A DM halo density gradient inside the stellar cluster would result in a broader MS, turn-off and RGB regions. This effect is usually attributed to photometric errors, variable reddening \citep{art-Carraroetal2002MNRAS}, extended star formation \citep{art-Twarogetal2011ApJL} and binaries \citep{art-Zhao2005AJ}. In the case of nuclear star clusters it could also be associated with the annihilation of DM particles inside the stars, given that within the typical size of nuclear clusters the DM density is expected to vary several orders of magnitude depending on the proximity of the galactic center.

For stellar clusters embedded in halos with extremely high DM densities we found an additional very strong signature: the bottom of the computed isochrones in the H-R diagram rises to higher luminosities because the low-mass stars, powered only with energy from DM annihilation, inflate and become fully convective. As this signature is hardly explained by other processes, we argue that this could be an indirect way to probe the presence of DM particles in the location of a cluster of stars.

\acknowledgments
We are grateful to the authors of DarkSUSY, from which some of the publicly available routines were adapted, and CESAM, as well as to the anonymous referee for a careful review. This work was supported by grants from FCT-MCTES (SFRH/BD/44321/2008) and Funda\c c\~ao Calouste Gulbenkian.

\newpage
\appendix
\section{Isochrone tables}
In Table \ref{tab-alliso} is shown a summary of the data used in Figures \ref{fig-isoe09} and \ref{fig-isoe10}, which corresponds to the isochrones of a classical stellar cluster and of stellar clusters embedded in halos of DM particles with densities $\rho_{\chi}=10^{9}\;$GeV cm$^{-3}$ and $\rho_{\chi}=10^{10}\;$GeV cm$^{-3}$. The mass of the stars ranges from 0.7 to 3.5 M$_{\odot}$ and their metallicity is Z=0.019. Our results do not rely on any specific initial mass function (IMF), i.e. any IMF could be used along with the table to obtain the relative number of stars in different sections of the isochrones.\\ 
\begin{longtable}{rrrrr}
\caption{Isochrones of a stellar cluster embedded in halos of DM particles with different DM densities. The DM particles are assumed to have a mass of 100 GeV and a spin-dependent scattering cross section with protons $\sigma_{\chi,SD}=10^{-38}\;$cm$^2$.}\\
$\rho_{\chi}$ (GeV cm$^{-3}$) & $Age$ (Myr) & $M ($M$_{\odot}$) & $\log ( T_{eff}$ / K ) & $\log ( L_{\star}$ / L$_{\odot}$ ) \\ \hline 
\endfirsthead

\multicolumn{5}{c}%
{{\tablename\ \thetable{} -- continued from previous page}} \\
$\rho_{\chi}$ (GeV cm$^{-3}$) & $Age$ (Myr) & $M ($M$_{\odot}$) & $\log ( T_{eff}$ / K ) & $\log ( L_{\star}$ / L$_{\odot}$ ) \\ \hline 
\endhead

\hline \multicolumn{5}{r}{{Continued on next page}} \\
\endfoot

\hline \hline
\endlastfoot

0 & 25 & 0.75000 & 3.67148 & -0.73140 \\
0 & 25 & 0.75000 & 3.67148 & -0.73140 \\ 
0 & 25 & 0.85000 & 3.70861 & -0.47804 \\ 
0 & 25 & 0.90000 & 3.72415 & -0.36338 \\ 
0 & 25 & 0.95000 & 3.73803 & -0.25484 \\ 
0 & 25 & 1.00000 & 3.75052 & -0.15104 \\ 
0 & 25 & 1.05000 & 3.76194 & -0.05041 \\ 
0 & 25 & 1.10000 & 3.77265 & 0.04843 \\ 
0 & 25 & 1.20000 & 3.79237 & 0.23570 \\ 
0 & 25 & 1.30000 & 3.81074 & 0.40377 \\ 
0 & 25 & 1.40000 & 3.82962 & 0.55440 \\ 
0 & 25 & 1.50000 & 3.85267 & 0.68920 \\ 
0 & 25 & 1.60000 & 3.87884 & 0.81074 \\ 
0 & 25 & 1.70000 & 3.90249 & 0.92201 \\ 
0 & 25 & 1.80000 & 3.92363 & 1.02498 \\ 
0 & 25 & 1.90000 & 3.94274 & 1.12120 \\ 
0 & 25 & 2.00000 & 3.96023 & 1.21166 \\ 
0 & 25 & 2.10000 & 3.97637 & 1.29708 \\ 
0 & 25 & 2.20000 & 3.99138 & 1.37803 \\ 
0 & 25 & 2.30000 & 4.00538 & 1.45504 \\ 
0 & 25 & 2.40000 & 4.01853 & 1.52842 \\ 
0 & 25 & 2.50000 & 4.03089 & 1.59858 \\
0 & 25 & 2.60000 & 4.04258 & 1.66576 \\ 
0 & 25 & 2.70000 & 4.05366 & 1.73024 \\ 
0 & 25 & 2.80000 & 4.06423 & 1.79230 \\ 
0 & 25 & 2.90000 & 4.07430 & 1.85203 \\ 
0 & 25 & 3.00000 & 4.08393 & 1.90969 \\ 
0 & 25 & 3.10000 & 4.09316 & 1.96535 \\ 
0 & 25 & 3.20000 & 4.10202 & 2.01921 \\ 
 & &&&\\
0 & 100 & 0.75000 & 3.66983 & -0.72340 \\ 
0 & 100 & 0.85000 & 3.70742 & -0.47299 \\ 
0 & 100 & 0.90000 & 3.72360 & -0.35737 \\ 
0 & 100 & 0.95000 & 3.73832 & -0.24661 \\ 
0 & 100 & 1.00000 & 3.75162 & -0.14020 \\ 
0 & 100 & 1.05000 & 3.76357 & -0.03808 \\ 
0 & 100 & 1.10000 & 3.77436 & 0.05996 \\ 
0 & 100 & 1.20000 & 3.79377 & 0.24417 \\ 
0 & 100 & 1.30000 & 3.81202 & 0.41114 \\ 
0 & 100 & 1.40000 & 3.83076 & 0.56072 \\ 
0 & 100 & 1.50000 & 3.85351 & 0.69513 \\ 
0 & 100 & 1.60000 & 3.87902 & 0.81715 \\ 
0 & 100 & 1.70000 & 3.90196 & 0.92923 \\ 
0 & 100 & 1.80000 & 3.92235 & 1.03330 \\ 
0 & 100 & 1.90000 & 3.94060 & 1.13064 \\ 
0 & 100 & 2.00000 & 3.95705 & 1.22230 \\ 
0 & 100 & 2.10000 & 3.97206 & 1.30915 \\ 
0 & 100 & 2.20000 & 3.98581 & 1.39169 \\ 
0 & 100 & 2.30000 & 3.99862 & 1.47068 \\ 
0 & 100 & 2.40000 & 4.01057 & 1.54636 \\ 
0 & 100 & 2.50000 & 4.02180 & 1.61927 \\ 
0 & 100 & 2.60000 & 4.03235 & 1.68947 \\ 
0 & 100 & 2.70000 & 4.04228 & 1.75733 \\ 
0 & 100 & 2.80000 & 4.05165 & 1.82309 \\ 
0 & 100 & 2.90000 & 4.06046 & 1.88690 \\ 
0 & 100 & 3.10000 & 4.07652 & 2.00952 \\ 
0 & 100 & 3.30000 & 4.09055 & 2.12621 \\ 
0 & 100 & 3.50000 & 4.10250 & 2.23821 \\ 
 & &&&\\
0 & 250 & 0.75000 & 3.66912 & -0.72153 \\ 
0 & 250 & 0.85000 & 3.70739 & -0.46915 \\ 
0 & 250 & 0.90000 & 3.72389 & -0.35231 \\ 
0 & 250 & 0.95000 & 3.73879 & -0.24068 \\ 
0 & 250 & 1.00000 & 3.75213 & -0.13382 \\ 
0 & 250 & 1.10000 & 3.77490 & 0.06741 \\ 
0 & 250 & 1.20000 & 3.79430 & 0.25283 \\ 
0 & 250 & 1.30000 & 3.81246 & 0.42100 \\ 
0 & 250 & 1.40000 & 3.83101 & 0.57182 \\ 
0 & 250 & 1.50000 & 3.85331 & 0.70756 \\ 
0 & 250 & 1.60000 & 3.87778 & 0.83108 \\ 
0 & 250 & 1.70000 & 3.89929 & 0.94477 \\ 
0 & 250 & 1.80000 & 3.91773 & 1.05040 \\ 
0 & 250 & 1.90000 & 3.93384 & 1.15023 \\ 
0 & 250 & 1.91000 & 3.93532 & 1.15989 \\ 
0 & 250 & 2.00000 & 3.94803 & 1.24461 \\ 
0 & 250 & 2.10000 & 3.96078 & 1.33514 \\ 
0 & 250 & 2.20000 & 3.97224 & 1.42220 \\ 
0 & 250 & 2.30000 & 3.98254 & 1.50639 \\ 
0 & 250 & 2.40000 & 3.99160 & 1.58793 \\ 
0 & 250 & 2.50000 & 3.99942 & 1.66736 \\ 
0 & 250 & 2.60000 & 4.00586 & 1.74458 \\ 
0 & 250 & 2.70000 & 4.01069 & 1.82007 \\ 
0 & 250 & 2.80000 & 4.01363 & 1.89408 \\ 
0 & 250 & 2.90000 & 4.01403 & 1.96626 \\ 
0 & 250 & 3.00000 & 4.01103 & 2.03676 \\ 
0 & 250 & 3.10000 & 4.00409 & 2.10587 \\ 
0 & 250 & 3.15000 & 4.00024 & 2.14164 \\ 
0 & 250 & 3.20000 & 4.01325 & 2.19874 \\ 
0 & 250 & 3.21000 & 4.04108 & 2.26802 \\ 
0 & 250 & 3.22000 & 4.02978 & 2.29505 \\ 
0 & 250 & 3.23000 & 4.02127 & 2.30979 \\ 
0 & 250 & 3.24000 & 4.00720 & 2.32118 \\ 
0 & 250 & 3.24500 & 3.98751 & 2.32297 \\ 
0 & 250 & 3.25000 & 3.95206 & 2.30835 \\ 
0 & 250 & 3.25150 & 3.93486 & 2.29574 \\ 
0 & 250 & 3.25350 & 3.90164 & 2.26602 \\ 
0 & 250 & 3.25500 & 3.89325 & 2.25815 \\ 
0 & 250 & 3.25590 & 3.87516 & 2.24086 \\ 
0 & 250 & 3.25600 & 3.83402 & 2.18768 \\ 
0 & 250 & 3.25630 & 3.82425 & 2.17807 \\ 
0 & 250 & 3.25650 & 3.81795 & 2.16721 \\ 
0 & 250 & 3.25680 & 3.78730 & 2.12451 \\ 
0 & 250 & 3.25710 & 3.78140 & 2.11338 \\ 
0 & 250 & 3.25750 & 3.77872 & 2.10775 \\ 
0 & 250 & 3.25800 & 3.76515 & 2.07537 \\ 
0 & 250 & 3.25830 & 3.75982 & 2.05500 \\ 
0 & 250 & 3.25870 & 3.74219 & 1.96290 \\ 
0 & 250 & 3.25900 & 3.73576 & 1.92345 \\ 
0 & 250 & 3.25950 & 3.72738 & 1.87437 \\ 
0 & 250 & 3.25980 & 3.71750 & 1.84125 \\ 
0 & 250 & 3.26000 & 3.71397 & 1.84491 \\ 
0 & 250 & 3.26100 & 3.70422 & 1.89534 \\ 
0 & 250 & 3.26200 & 3.69260 & 2.01970 \\ 
0 & 250 & 3.26300 & 3.68374 & 2.13944 \\ 
0 & 250 & 3.26500 & 3.67265 & 2.29575 \\ 
0 & 250 & 3.26700 & 3.66270 & 2.43925 \\ 
 & &&&\\
0 & 500 & 0.75000 & 3.66911 & -0.71835 \\ 
0 & 500 & 0.85000 & 3.70780 & -0.46408 \\ 
0 & 500 & 0.90000 & 3.72443 & -0.34634 \\ 
0 & 500 & 0.95000 & 3.73938 & -0.23390 \\ 
0 & 500 & 1.00000 & 3.75276 & -0.12610 \\ 
0 & 500 & 1.10000 & 3.77554 & 0.07749 \\ 
0 & 500 & 1.20000 & 3.79490 & 0.26571 \\ 
0 & 500 & 1.30000 & 3.81286 & 0.43662 \\ 
0 & 500 & 1.40000 & 3.83086 & 0.58981 \\ 
0 & 500 & 1.50000 & 3.85171 & 0.72776 \\ 
0 & 500 & 1.60000 & 3.87358 & 0.85341 \\ 
0 & 500 & 1.70000 & 3.89155 & 0.96972 \\ 
0 & 500 & 1.80000 & 3.90572 & 1.07833 \\ 
0 & 500 & 1.90000 & 3.91734 & 1.18272 \\ 
0 & 500 & 2.00000 & 3.92637 & 1.28295 \\ 
0 & 500 & 2.04000 & 3.92920 & 1.32207 \\ 
0 & 500 & 2.10000 & 3.93249 & 1.37987 \\ 
0 & 500 & 2.20000 & 3.93491 & 1.47332 \\ 
0 & 500 & 2.30000 & 3.93224 & 1.56312 \\ 
0 & 500 & 2.40000 & 3.92290 & 1.64869 \\ 
0 & 500 & 2.46000 & 3.92378 & 1.71275 \\ 
0 & 500 & 2.46500 & 3.92800 & 1.72412 \\ 
0 & 500 & 2.47000 & 3.93591 & 1.74035 \\ 
0 & 500 & 2.47500 & 3.95797 & 1.78130 \\ 
0 & 500 & 2.48000 & 3.95911 & 1.82439 \\ 
0 & 500 & 2.50000 & 3.93963 & 1.86087 \\ 
0 & 500 & 2.51000 & 3.91183 & 1.86722 \\ 
0 & 500 & 2.51200 & 3.90364 & 1.86503 \\ 
0 & 500 & 2.51400 & 3.88457 & 1.85568 \\ 
0 & 500 & 2.51600 & 3.86064 & 1.83821 \\ 
0 & 500 & 2.51800 & 3.82350 & 1.80095 \\ 
0 & 500 & 2.51900 & 3.80242 & 1.77489 \\ 
0 & 500 & 2.52000 & 3.78050 & 1.73315 \\ 
0 & 500 & 2.52100 & 3.76165 & 1.65732 \\ 
0 & 500 & 2.52200 & 3.74752 & 1.55854 \\ 
0 & 500 & 2.52300 & 3.73554 & 1.48007 \\ 
0 & 500 & 2.52400 & 3.72477 & 1.43067 \\ 
0 & 500 & 2.52500 & 3.71456 & 1.45750 \\ 
0 & 500 & 2.52700 & 3.69870 & 1.65553 \\ 
0 & 500 & 2.53000 & 3.68516 & 1.87425 \\ 
0 & 500 & 2.53500 & 3.66891 & 2.12896 \\ 
0 & 500 & 2.53700 & 3.66378 & 2.20523 \\ 
0 & 500 & 2.53810 & 3.66086 & 2.24817 \\ 
 & &&&\\
0 & 1000 & 0.75000 & 3.66966 & -0.71245 \\ 
0 & 1000 & 0.85000 & 3.70876 & -0.45537 \\ 
0 & 1000 & 0.90000 & 3.72551 & -0.33606 \\ 
0 & 1000 & 0.95000 & 3.74054 & -0.22176 \\ 
0 & 1000 & 1.00000 & 3.75396 & -0.11176 \\ 
0 & 1000 & 1.10000 & 3.77672 & 0.09725 \\ 
0 & 1000 & 1.20000 & 3.79591 & 0.29159 \\ 
0 & 1000 & 1.30000 & 3.81316 & 0.46803 \\ 
0 & 1000 & 1.40000 & 3.82872 & 0.62433 \\ 
0 & 1000 & 1.45000 & 3.83611 & 0.69628 \\ 
0 & 1000 & 1.50000 & 3.84341 & 0.76530 \\ 
0 & 1000 & 1.54000 & 3.84855 & 0.81728 \\ 
0 & 1000 & 1.60000 & 3.85482 & 0.89291 \\ 
0 & 1000 & 1.70000 & 3.86015 & 1.01260 \\ 
0 & 1000 & 1.80000 & 3.85713 & 1.12389 \\ 
0 & 1000 & 1.90000 & 3.84588 & 1.23260 \\ 
0 & 1000 & 1.91000 & 3.84767 & 1.24875 \\ 
0 & 1000 & 1.91500 & 3.85008 & 1.25887 \\ 
0 & 1000 & 1.92000 & 3.85709 & 1.27678 \\ 
0 & 1000 & 1.92500 & 3.87040 & 1.30607 \\ 
0 & 1000 & 1.93000 & 3.88797 & 1.37559 \\ 
0 & 1000 & 1.95000 & 3.87245 & 1.41807 \\ 
0 & 1000 & 1.95400 & 3.86743 & 1.42494 \\ 
0 & 1000 & 1.96200 & 3.84673 & 1.43204 \\ 
0 & 1000 & 1.96600 & 3.83163 & 1.42992 \\ 
0 & 1000 & 1.96800 & 3.81836 & 1.42292 \\ 
0 & 1000 & 1.96900 & 3.80306 & 1.40808 \\ 
0 & 1000 & 1.97000 & 3.79564 & 1.39787 \\ 
0 & 1000 & 1.97400 & 3.76855 & 1.32206 \\ 
0 & 1000 & 1.97600 & 3.74509 & 1.19045 \\ 
0 & 1000 & 1.97800 & 3.72258 & 1.08958 \\ 
0 & 1000 & 1.97900 & 3.71797 & 1.09934 \\ 
0 & 1000 & 1.98100 & 3.71056 & 1.15725 \\ 
0 & 1000 & 1.98200 & 3.70709 & 1.20857 \\ 
0 & 1000 & 1.99000 & 3.69320 & 1.48019 \\ 
0 & 1000 & 2.00000 & 3.67947 & 1.73567 \\ 
0 & 1000 & 2.02000 & 3.65704 & 2.11526 \\ 
0 & 1000 & 2.03000 & 3.61677 & 2.69923 \\
 & &&&\\
0 & 2500 & 0.70000 & 3.64888 & -0.83902 \\ 
0 & 2500 & 0.75000 & 3.67183 & -0.69615 \\ 
0 & 2500 & 0.85000 & 3.71169 & -0.43088 \\ 
0 & 2500 & 0.90000 & 3.72873 & -0.30623 \\ 
0 & 2500 & 0.95000 & 3.74393 & -0.18563 \\ 
0 & 2500 & 1.00000 & 3.75735 & -0.06825 \\ 
0 & 2500 & 1.10000 & 3.77973 & 0.15955 \\ 
0 & 2500 & 1.20000 & 3.79652 & 0.37132 \\ 
0 & 2500 & 1.25000 & 3.80143 & 0.46383 \\ 
0 & 2500 & 1.30000 & 3.80298 & 0.54658 \\ 
0 & 2500 & 1.36000 & 3.79974 & 0.63472 \\ 
0 & 2500 & 1.41000 & 3.79967 & 0.72635 \\ 
0 & 2500 & 1.41300 & 3.80651 & 0.76339 \\ 
0 & 2500 & 1.41600 & 3.81580 & 0.81974 \\ 
0 & 2500 & 1.42000 & 3.81431 & 0.83524 \\ 
0 & 2500 & 1.43000 & 3.81240 & 0.85975 \\ 
0 & 2500 & 1.44000 & 3.80895 & 0.88391 \\ 
0 & 2500 & 1.45000 & 3.80463 & 0.90585 \\ 
0 & 2500 & 1.46000 & 3.79809 & 0.92650 \\ 
0 & 2500 & 1.46400 & 3.79489 & 0.93310 \\ 
0 & 2500 & 1.46800 & 3.78899 & 0.93793 \\ 
0 & 2500 & 1.47200 & 3.78720 & 0.94176 \\ 
0 & 2500 & 1.47600 & 3.77708 & 0.93353 \\ 
0 & 2500 & 1.48000 & 3.76533 & 0.90942 \\ 
0 & 2500 & 1.48400 & 3.74951 & 0.85217 \\ 
0 & 2500 & 1.48500 & 3.75394 & 0.87314 \\ 
0 & 2500 & 1.48800 & 3.73419 & 0.78814 \\ 
0 & 2500 & 1.49000 & 3.72322 & 0.75467 \\ 
0 & 2500 & 1.49200 & 3.72011 & 0.75224 \\ 
0 & 2500 & 1.49600 & 3.71147 & 0.77299 \\ 
0 & 2500 & 1.50000 & 3.70672 & 0.82157 \\ 
0 & 2500 & 1.51000 & 3.70153 & 0.93898 \\ 
0 & 2500 & 1.52000 & 3.69624 & 1.09652 \\ 
0 & 2500 & 1.53000 & 3.68671 & 1.33581 \\ 
0 & 2500 & 1.53500 & 3.67794 & 1.51972 \\ 
0 & 2500 & 1.53800 & 3.67202 & 1.63355 \\ 
0 & 2500 & 1.53970 & 3.66811 & 1.70777 \\ 
 & &&&\\
0 & 5000 & 0.70000 & 3.65207 & -0.81780 \\ 
0 & 5000 & 0.75000 & 3.67569 & -0.66929 \\ 
0 & 5000 & 0.85000 & 3.71673 & -0.38839 \\ 
0 & 5000 & 0.90000 & 3.73415 & -0.25306 \\ 
0 & 5000 & 0.95000 & 3.74950 & -0.11874 \\ 
0 & 5000 & 1.00000 & 3.76244 & 0.01398 \\ 
0 & 5000 & 1.05000 & 3.77306 & 0.14705 \\ 
0 & 5000 & 1.10000 & 3.78020 & 0.27460 \\ 
0 & 5000 & 1.14000 & 3.78397 & 0.38615 \\ 
0 & 5000 & 1.15000 & 3.78457 & 0.41565 \\ 
0 & 5000 & 1.17000 & 3.78514 & 0.47905 \\ 
0 & 5000 & 1.18000 & 3.78441 & 0.51244 \\ 
0 & 5000 & 1.19000 & 3.78335 & 0.54390 \\ 
0 & 5000 & 1.20000 & 3.78154 & 0.57640 \\ 
0 & 5000 & 1.21000 & 3.77751 & 0.61223 \\ 
0 & 5000 & 1.22000 & 3.77125 & 0.64143 \\ 
0 & 5000 & 1.22400 & 3.76601 & 0.64974 \\ 
0 & 5000 & 1.22800 & 3.75923 & 0.65024 \\ 
0 & 5000 & 1.23200 & 3.74906 & 0.63586 \\ 
0 & 5000 & 1.23300 & 3.74200 & 0.61838 \\ 
0 & 5000 & 1.23400 & 3.74069 & 0.61596 \\ 
0 & 5000 & 1.23500 & 3.74155 & 0.61948 \\ 
0 & 5000 & 1.23600 & 3.72723 & 0.58247 \\ 
0 & 5000 & 1.23700 & 3.72369 & 0.57655 \\ 
0 & 5000 & 1.23900 & 3.71812 & 0.57240 \\ 
0 & 5000 & 1.24000 & 3.71011 & 0.58288 \\ 
0 & 5000 & 1.24600 & 3.70190 & 0.65929 \\ 
0 & 5000 & 1.24800 & 3.70018 & 0.69494 \\ 
0 & 5000 & 1.24900 & 3.69792 & 0.75628 \\ 
0 & 5000 & 1.25000 & 3.69732 & 0.77741 \\ 
0 & 5000 & 1.25500 & 3.69445 & 0.88494 \\ 
0 & 5000 & 1.26000 & 3.68739 & 1.10785 \\ 
0 & 5000 & 1.27000 & 3.65948 & 1.71237 \\ 
0 & 5000 & 1.27100 & 3.65081 & 1.86220 \\ 
 & &&&\\
0 & 10000 & 0.70000 & 3.65883 & -0.77354 \\ 
0 & 10000 & 0.75000 & 3.68387 & -0.61118 \\ 
0 & 10000 & 0.85000 & 3.72710 & -0.28840 \\ 
0 & 10000 & 0.90000 & 3.74406 & -0.12341 \\ 
0 & 10000 & 0.95000 & 3.75684 & 0.05096 \\ 
0 & 10000 & 0.96000 & 3.75849 & 0.08575 \\ 
0 & 10000 & 0.97000 & 3.75991 & 0.12365 \\ 
0 & 10000 & 0.99000 & 3.76159 & 0.20510 \\ 
0 & 10000 & 0.99600 & 3.76163 & 0.23140 \\ 
0 & 10000 & 1.00300 & 3.76118 & 0.26305 \\ 
0 & 10000 & 1.00800 & 3.76031 & 0.28752 \\ 
0 & 10000 & 1.01200 & 3.75935 & 0.30657 \\ 
0 & 10000 & 1.01800 & 3.75698 & 0.33548 \\ 
0 & 10000 & 1.02200 & 3.75448 & 0.35441 \\ 
0 & 10000 & 1.02700 & 3.74890 & 0.37717 \\ 
0 & 10000 & 1.03000 & 3.74360 & 0.38685 \\ 
0 & 10000 & 1.03200 & 3.73903 & 0.39007 \\ 
0 & 10000 & 1.03400 & 3.73225 & 0.38940 \\ 
0 & 10000 & 1.03500 & 3.72750 & 0.38722 \\ 
0 & 10000 & 1.03600 & 3.72339 & 0.38579 \\ 
0 & 10000 & 1.03680 & 3.71901 & 0.38531 \\ 
0 & 10000 & 1.03720 & 3.71672 & 0.38593 \\ 
0 & 10000 & 1.03800 & 3.71246 & 0.38952 \\ 
0 & 10000 & 1.03880 & 3.70827 & 0.39788 \\ 
0 & 10000 & 1.04000 & 3.70379 & 0.41656 \\ 
0 & 10000 & 1.04030 & 3.70263 & 0.42389 \\ 
0 & 10000 & 1.04050 & 3.70193 & 0.42897 \\ 
0 & 10000 & 1.04070 & 3.70126 & 0.43440 \\ 
0 & 10000 & 1.04100 & 3.70031 & 0.44326 \\ 
0 & 10000 & 1.04130 & 3.69943 & 0.45274 \\ 
0 & 10000 & 1.04160 & 3.69878 & 0.46080 \\ 
0 & 10000 & 1.04190 & 3.69793 & 0.47273 \\ 
0 & 10000 & 1.04300 & 3.69567 & 0.51601 \\ 
0 & 10000 & 1.04380 & 3.69436 & 0.55155 \\ 
0 & 10000 & 1.04500 & 3.69276 & 0.61013 \\ 
0 & 10000 & 1.05000 & 3.68530 & 0.92676 \\ 
0 & 10000 & 1.05500 & 3.65768 & 1.59461 \\ 
0 & 10000 & 1.05600 & 3.63176 & 2.03457 \\
& & & & \\
$10^{9}$ & 25 & 0.75000 & 3.65940 & -0.77514 \\ 
$10^{9}$ & 25 & 0.80000 & 3.67935 & -0.64364 \\ 
$10^{9}$ & 25 & 0.90000 & 3.71323 & -0.40504 \\ 
$10^{9}$ & 25 & 1.00000 & 3.74020 & -0.19188 \\ 
$10^{9}$ & 25 & 1.10000 & 3.76673 & 0.03980 \\ 
$10^{9}$ & 25 & 1.20000 & 3.78723 & 0.22969 \\ 
$10^{9}$ & 25 & 1.30000 & 3.80465 & 0.37837 \\ 
$10^{9}$ & 25 & 1.40000 & 3.82367 & 0.53444 \\ 
$10^{9}$ & 25 & 1.45000 & 3.83408 & 0.60678 \\ 
$10^{9}$ & 25 & 1.50000 & 3.84579 & 0.67480 \\ 
$10^{9}$ & 25 & 1.55000 & 3.85956 & 0.74008 \\ 
$10^{9}$ & 25 & 1.60000 & 3.87333 & 0.80141 \\ 
$10^{9}$ & 25 & 1.70000 & 3.89843 & 0.91551 \\ 
$10^{9}$ & 25 & 1.80000 & 3.92057 & 1.02029 \\ 
$10^{9}$ & 25 & 1.90000 & 3.94038 & 1.11764 \\ 
$10^{9}$ & 25 & 2.00000 & 3.95837 & 1.20887 \\ 
$10^{9}$ & 25 & 2.05000 & 3.96680 & 1.25246 \\ 
$10^{9}$ & 25 & 2.20000 & 3.99024 & 1.37628 \\ 
$10^{9}$ & 25 & 2.30000 & 4.00447 & 1.45354 \\ 
$10^{9}$ & 25 & 2.50300 & 4.03062 & 1.59952 \\ 
$10^{9}$ & 25 & 2.70000 & 4.05324 & 1.72939 \\ 
$10^{9}$ & 25 & 2.90000 & 4.07398 & 1.85132 \\ 
$10^{9}$ & 25 & 3.10000 & 4.09295 & 1.96473 \\ 
$10^{9}$ & 25 & 3.20000 & 4.10184 & 2.01861 \\ 
 & &&&\\
$10^{9}$ & 100 & 0.75000 & 3.64806 & -0.78361 \\ 
$10^{9}$ & 100 & 0.80000 & 3.67539 & -0.63812 \\ 
$10^{9}$ & 100 & 0.90000 & 3.70817 & -0.41027 \\ 
$10^{9}$ & 100 & 1.00000 & 3.73923 & -0.18932 \\ 
$10^{9}$ & 100 & 1.10000 & 3.76768 & 0.04934 \\ 
$10^{9}$ & 100 & 1.20000 & 3.78884 & 0.23968 \\ 
$10^{9}$ & 100 & 1.30000 & 3.80555 & 0.38429 \\ 
$10^{9}$ & 100 & 1.40000 & 3.82488 & 0.54176 \\ 
$10^{9}$ & 100 & 1.50000 & 3.84740 & 0.68206 \\ 
$10^{9}$ & 100 & 1.55000 & 3.86078 & 0.74663 \\ 
$10^{9}$ & 100 & 1.60000 & 3.87416 & 0.80796 \\ 
$10^{9}$ & 100 & 1.70000 & 3.89848 & 0.92257 \\ 
$10^{9}$ & 100 & 1.80000 & 3.91986 & 1.02830 \\ 
$10^{9}$ & 100 & 1.85000 & 3.92962 & 1.07836 \\ 
$10^{9}$ & 100 & 1.95000 & 3.94760 & 1.17383 \\ 
$10^{9}$ & 100 & 2.00000 & 3.95586 & 1.21921 \\ 
$10^{9}$ & 100 & 2.10000 & 3.97120 & 1.30657 \\ 
$10^{9}$ & 100 & 2.20000 & 3.98528 & 1.38950 \\ 
$10^{9}$ & 100 & 2.30000 & 3.99825 & 1.46867 \\ 
$10^{9}$ & 100 & 2.40000 & 4.01033 & 1.54459 \\ 
$10^{9}$ & 100 & 2.55000 & 4.02706 & 1.65298 \\ 
$10^{9}$ & 100 & 2.70000 & 4.04232 & 1.75583 \\ 
$10^{9}$ & 100 & 2.90000 & 4.06061 & 1.88549 \\ 
$10^{9}$ & 100 & 3.10000 & 4.07678 & 2.00799 \\ 
$10^{9}$ & 100 & 3.30000 & 4.09091 & 2.12462 \\ 
$10^{9}$ & 100 & 3.50000 & 4.10297 & 2.23652 \\ 
 & &&&\\
$10^{9}$ & 250 & 0.70000 & 3.62785 & -0.90881 \\ 
$10^{9}$ & 250 & 0.75000 & 3.64613 & -0.78529 \\ 
$10^{9}$ & 250 & 0.80000 & 3.66922 & -0.65203 \\ 
$10^{9}$ & 250 & 0.85000 & 3.68911 & -0.52894 \\ 
$10^{9}$ & 250 & 0.90000 & 3.70751 & -0.40976 \\ 
$10^{9}$ & 250 & 0.95000 & 3.72422 & -0.29562 \\ 
$10^{9}$ & 250 & 1.00000 & 3.73964 & -0.18478 \\ 
$10^{9}$ & 250 & 1.10000 & 3.76808 & 0.05528 \\ 
$10^{9}$ & 250 & 1.20000 & 3.78943 & 0.24738 \\ 
$10^{9}$ & 250 & 1.30000 & 3.80616 & 0.39328 \\ 
$10^{9}$ & 250 & 1.40000 & 3.82525 & 0.55156 \\ 
$10^{9}$ & 250 & 1.50000 & 3.84766 & 0.69357 \\ 
$10^{9}$ & 250 & 1.60000 & 3.87383 & 0.82152 \\ 
$10^{9}$ & 250 & 1.70000 & 3.89678 & 0.93757 \\ 
$10^{9}$ & 250 & 1.80000 & 3.91656 & 1.04530 \\ 
$10^{9}$ & 250 & 1.90000 & 3.93334 & 1.14580 \\ 
$10^{9}$ & 250 & 2.00000 & 3.94805 & 1.24091 \\ 
$10^{9}$ & 250 & 2.10000 & 3.96105 & 1.33161 \\ 
$10^{9}$ & 250 & 2.20000 & 3.97272 & 1.41878 \\ 
$10^{9}$ & 250 & 2.30000 & 3.98325 & 1.50318 \\ 
$10^{9}$ & 250 & 2.40000 & 3.99248 & 1.58492 \\ 
$10^{9}$ & 250 & 2.50000 & 4.00055 & 1.66414 \\ 
$10^{9}$ & 250 & 2.60000 & 4.00723 & 1.74164 \\ 
$10^{9}$ & 250 & 2.70000 & 4.01235 & 1.81718 \\ 
$10^{9}$ & 250 & 2.80000 & 4.01568 & 1.89125 \\ 
$10^{9}$ & 250 & 2.90000 & 4.01648 & 1.96371 \\ 
$10^{9}$ & 250 & 3.00000 & 4.01418 & 2.03443 \\ 
$10^{9}$ & 250 & 3.10000 & 4.00774 & 2.10327 \\ 
$10^{9}$ & 250 & 3.20000 & 4.00364 & 2.18023 \\ 
$10^{9}$ & 250 & 3.22400 & 4.03716 & 2.24556 \\ 
$10^{9}$ & 250 & 3.22500 & 4.04565 & 2.26296 \\ 
$10^{9}$ & 250 & 3.22700 & 4.04611 & 2.27168 \\ 
$10^{9}$ & 250 & 3.24000 & 4.03188 & 2.30772 \\ 
$10^{9}$ & 250 & 3.24700 & 4.02709 & 2.31824 \\ 
$10^{9}$ & 250 & 3.26000 & 4.00662 & 2.33630 \\ 
$10^{9}$ & 250 & 3.26400 & 3.99822 & 2.33867 \\ 
$10^{9}$ & 250 & 3.26800 & 3.97859 & 2.33592 \\ 
$10^{9}$ & 250 & 3.27200 & 3.95230 & 2.32383 \\ 
$10^{9}$ & 250 & 3.27300 & 3.91576 & 2.29236 \\ 
$10^{9}$ & 250 & 3.27400 & 3.90395 & 2.28243 \\ 
$10^{9}$ & 250 & 3.27500 & 3.87751 & 2.25651 \\ 
$10^{9}$ & 250 & 3.27600 & 3.86864 & 2.24285 \\ 
$10^{9}$ & 250 & 3.27700 & 3.83463 & 2.20543 \\ 
$10^{9}$ & 250 & 3.27730 & 3.82712 & 2.19689 \\ 
$10^{9}$ & 250 & 3.27740 & 3.82298 & 2.19149 \\ 
$10^{9}$ & 250 & 3.27745 & 3.82002 & 2.18525 \\ 
$10^{9}$ & 250 & 3.27747 & 3.82052 & 2.18883 \\ 
$10^{9}$ & 250 & 3.27749 & 3.77211 & 2.10763 \\ 
$10^{9}$ & 250 & 3.27750 & 3.77179 & 2.10691 \\ 
$10^{9}$ & 250 & 3.27755 & 3.77195 & 2.10634 \\ 
$10^{9}$ & 250 & 3.27760 & 3.76892 & 2.09917 \\ 
$10^{9}$ & 250 & 3.27780 & 3.76524 & 2.09003 \\ 
$10^{9}$ & 250 & 3.27800 & 3.75722 & 2.05817 \\ 
$10^{9}$ & 250 & 3.27900 & 3.76627 & 2.09479 \\ 
$10^{9}$ & 250 & 3.28000 & 3.74096 & 1.97505 \\ 
$10^{9}$ & 250 & 3.28020 & 3.73469 & 1.93495 \\ 
$10^{9}$ & 250 & 3.28040 & 3.73238 & 1.92099 \\ 
$10^{9}$ & 250 & 3.28080 & 3.72446 & 1.87712 \\ 
$10^{9}$ & 250 & 3.28085 & 3.72644 & 1.88832 \\ 
$10^{9}$ & 250 & 3.28090 & 3.72762 & 1.89205 \\ 
$10^{9}$ & 250 & 3.28095 & 3.72441 & 1.87706 \\ 
$10^{9}$ & 250 & 3.28097 & 3.72408 & 1.87363 \\ 
$10^{9}$ & 250 & 3.28098 & 3.71088 & 1.87058 \\ 
$10^{9}$ & 250 & 3.28099 & 3.70322 & 1.91200 \\ 
$10^{9}$ & 250 & 3.28100 & 3.70327 & 1.91151 \\ 
$10^{9}$ & 250 & 3.28200 & 3.69780 & 1.96676 \\ 
$10^{9}$ & 250 & 3.28250 & 3.68832 & 2.08389 \\ 
 & &&&\\
$10^{9}$ & 500 & 0.75000 & 3.64408 & -0.79070 \\ 
$10^{9}$ & 500 & 0.80000 & 3.66957 & -0.64842 \\ 
$10^{9}$ & 500 & 0.90000 & 3.70757 & -0.40717 \\ 
$10^{9}$ & 500 & 1.00000 & 3.74034 & -0.17884 \\ 
$10^{9}$ & 500 & 1.10000 & 3.76910 & 0.06393 \\ 
$10^{9}$ & 500 & 1.20000 & 3.79009 & 0.25818 \\ 
$10^{9}$ & 500 & 1.30000 & 3.80671 & 0.40618 \\ 
$10^{9}$ & 500 & 1.40000 & 3.82572 & 0.56822 \\ 
$10^{9}$ & 500 & 1.50000 & 3.84757 & 0.71313 \\ 
$10^{9}$ & 500 & 1.59000 & 3.86913 & 0.83031 \\ 
$10^{9}$ & 500 & 1.70000 & 3.89115 & 0.96164 \\ 
$10^{9}$ & 500 & 1.80000 & 3.90685 & 1.07248 \\ 
$10^{9}$ & 500 & 1.90000 & 3.91915 & 1.17711 \\ 
$10^{9}$ & 500 & 2.00000 & 3.92890 & 1.27779 \\ 
$10^{9}$ & 500 & 2.05000 & 3.93269 & 1.32681 \\ 
$10^{9}$ & 500 & 2.10000 & 3.93572 & 1.37506 \\ 
$10^{9}$ & 500 & 2.20000 & 3.93900 & 1.46899 \\ 
$10^{9}$ & 500 & 2.30000 & 3.93758 & 1.55986 \\ 
$10^{9}$ & 500 & 2.40000 & 3.92935 & 1.64569 \\ 
$10^{9}$ & 500 & 2.45000 & 3.92343 & 1.68908 \\ 
$10^{9}$ & 500 & 2.46000 & 3.92284 & 1.69862 \\ 
$10^{9}$ & 500 & 2.48000 & 3.92464 & 1.72265 \\ 
$10^{9}$ & 500 & 2.49000 & 3.93121 & 1.74291 \\ 
$10^{9}$ & 500 & 2.49300 & 3.93461 & 1.75012 \\ 
$10^{9}$ & 500 & 2.49700 & 3.94584 & 1.76993 \\ 
$10^{9}$ & 500 & 2.49850 & 3.95222 & 1.78142 \\ 
$10^{9}$ & 500 & 2.49900 & 3.95715 & 1.79112 \\ 
$10^{9}$ & 500 & 2.49960 & 3.96773 & 1.81300 \\ 
$10^{9}$ & 500 & 2.50300 & 3.96527 & 1.83684 \\ 
$10^{9}$ & 500 & 2.50600 & 3.96201 & 1.84632 \\ 
$10^{9}$ & 500 & 2.51000 & 3.95950 & 1.85424 \\ 
$10^{9}$ & 500 & 2.52000 & 3.95287 & 1.87321 \\ 
$10^{9}$ & 500 & 2.53000 & 3.94197 & 1.88995 \\ 
$10^{9}$ & 500 & 2.54000 & 3.91884 & 1.89836 \\ 
$10^{9}$ & 500 & 2.54200 & 3.90416 & 1.89429 \\ 
$10^{9}$ & 500 & 2.54600 & 3.86379 & 1.86872 \\ 
$10^{9}$ & 500 & 2.54700 & 3.85697 & 1.86313 \\ 
$10^{9}$ & 500 & 2.54900 & 3.81757 & 1.82244 \\ 
$10^{9}$ & 500 & 2.54930 & 3.80455 & 1.80574 \\ 
$10^{9}$ & 500 & 2.54960 & 3.80038 & 1.79991 \\ 
$10^{9}$ & 500 & 2.55000 & 3.76616 & 1.70926 \\ 
$10^{9}$ & 500 & 2.55100 & 3.75226 & 1.63132 \\ 
$10^{9}$ & 500 & 2.55150 & 3.75690 & 1.65548 \\ 
$10^{9}$ & 500 & 2.55200 & 3.75174 & 1.62251 \\ 
$10^{9}$ & 500 & 2.55250 & 3.74014 & 1.53703 \\ 
$10^{9}$ & 500 & 2.55260 & 3.73943 & 1.53622 \\ 
$10^{9}$ & 500 & 2.55280 & 3.73715 & 1.51841 \\ 
$10^{9}$ & 500 & 2.55285 & 3.73741 & 1.52336 \\ 
$10^{9}$ & 500 & 2.55290 & 3.72144 & 1.45710 \\ 
$10^{9}$ & 500 & 2.55295 & 3.72082 & 1.45723 \\ 
$10^{9}$ & 500 & 2.55300 & 3.72156 & 1.46155 \\ 
$10^{9}$ & 500 & 2.55350 & 3.71900 & 1.45872 \\ 
$10^{9}$ & 500 & 2.55400 & 3.71281 & 1.49296 \\ 
$10^{9}$ & 500 & 2.55500 & 3.70704 & 1.55170 \\ 
$10^{9}$ & 500 & 2.55600 & 3.70201 & 1.62113 \\ 
$10^{9}$ & 500 & 2.56000 & 3.68100 & 1.95350 \\ 
$10^{9}$ & 500 & 2.56100 & 3.67724 & 2.01236 \\ 
$10^{9}$ & 500 & 2.56200 & 3.66962 & 2.12881 \\ 
$10^{9}$ & 500 & 2.56300 & 3.66625 & 2.17883 \\ 
$10^{9}$ & 500 & 2.56350 & 3.66608 & 2.18152 \\ 
 & &&&\\
$10^{9}$ & 1000 & 0.75000 & 3.64863 & -0.77614 \\ 
$10^{9}$ & 1000 & 0.80000 & 3.66765 & -0.65340 \\ 
$10^{9}$ & 1000 & 0.85000 & 3.69086 & -0.51828 \\ 
$10^{9}$ & 1000 & 0.90000 & 3.70809 & -0.40172 \\ 
$10^{9}$ & 1000 & 0.95000 & 3.72571 & -0.28306 \\ 
$10^{9}$ & 1000 & 1.00000 & 3.74146 & -0.16867 \\ 
$10^{9}$ & 1000 & 1.05000 & 3.75528 & -0.05891 \\ 
$10^{9}$ & 1000 & 1.10000 & 3.77047 & 0.07985 \\ 
$10^{9}$ & 1000 & 1.15000 & 3.78125 & 0.18185 \\ 
$10^{9}$ & 1000 & 1.20000 & 3.79132 & 0.28015 \\ 
$10^{9}$ & 1000 & 1.25000 & 3.79846 & 0.34292 \\ 
$10^{9}$ & 1000 & 1.30000 & 3.80798 & 0.43478 \\ 
$10^{9}$ & 1000 & 1.36000 & 3.81870 & 0.53583 \\ 
$10^{9}$ & 1000 & 1.40000 & 3.82578 & 0.60156 \\ 
$10^{9}$ & 1000 & 1.45000 & 3.83472 & 0.67831 \\ 
$10^{9}$ & 1000 & 1.50000 & 3.84384 & 0.75070 \\ 
$10^{9}$ & 1000 & 1.55000 & 3.85216 & 0.81785 \\ 
$10^{9}$ & 1000 & 1.60000 & 3.85892 & 0.88337 \\ 
$10^{9}$ & 1000 & 1.65000 & 3.86407 & 0.94516 \\ 
$10^{9}$ & 1000 & 1.70000 & 3.86683 & 1.00491 \\ 
$10^{9}$ & 1000 & 1.75000 & 3.86835 & 1.06331 \\ 
$10^{9}$ & 1000 & 1.80000 & 3.86693 & 1.11966 \\ 
$10^{9}$ & 1000 & 1.85000 & 3.86274 & 1.17342 \\ 
$10^{9}$ & 1000 & 1.90000 & 3.85571 & 1.22530 \\ 
$10^{9}$ & 1000 & 1.93000 & 3.85144 & 1.25821 \\ 
$10^{9}$ & 1000 & 1.94000 & 3.85155 & 1.27213 \\ 
$10^{9}$ & 1000 & 1.95000 & 3.85502 & 1.29060 \\ 
$10^{9}$ & 1000 & 1.96000 & 3.87631 & 1.33767 \\ 
$10^{9}$ & 1000 & 1.97000 & 3.89319 & 1.41817 \\ 
$10^{9}$ & 1000 & 1.98000 & 3.88974 & 1.44101 \\ 
$10^{9}$ & 1000 & 1.99000 & 3.88288 & 1.46386 \\ 
$10^{9}$ & 1000 & 2.00000 & 3.87156 & 1.48272 \\ 
$10^{9}$ & 1000 & 2.00500 & 3.85922 & 1.48955 \\ 
$10^{9}$ & 1000 & 2.00800 & 3.84669 & 1.48975 \\ 
$10^{9}$ & 1000 & 2.01100 & 3.83247 & 1.48554 \\ 
$10^{9}$ & 1000 & 2.01400 & 3.80006 & 1.46040 \\ 
$10^{9}$ & 1000 & 2.01700 & 3.77624 & 1.40920 \\ 
$10^{9}$ & 1000 & 2.01800 & 3.76002 & 1.33634 \\ 
$10^{9}$ & 1000 & 2.02000 & 3.73630 & 1.18505 \\ 
$10^{9}$ & 1000 & 2.02050 & 3.73211 & 1.16281 \\ 
$10^{9}$ & 1000 & 2.02100 & 3.72376 & 1.13597 \\ 
$10^{9}$ & 1000 & 2.02150 & 3.71762 & 1.14516 \\ 
$10^{9}$ & 1000 & 2.02200 & 3.71595 & 1.15342 \\ 
$10^{9}$ & 1000 & 2.02300 & 3.71265 & 1.17846 \\ 
$10^{9}$ & 1000 & 2.02400 & 3.70511 & 1.28258 \\ 
$10^{9}$ & 1000 & 2.02800 & 3.69387 & 1.49529 \\ 
$10^{9}$ & 1000 & 2.03000 & 3.68954 & 1.57652 \\ 
$10^{9}$ & 1000 & 2.04000 & 3.67028 & 1.91074 \\ 
$10^{9}$ & 1000 & 2.05000 & 3.65965 & 2.08876 \\ 
$10^{9}$ & 1000 & 2.06000 & 3.62073 & 2.65588 \\ 
 & &&&\\
$10^{9}$ & 2500 & 0.75000 & 3.64424 & -0.78350 \\ 
$10^{9}$ & 2500 & 0.80000 & 3.66878 & -0.64369 \\ 
$10^{9}$ & 2500 & 0.85000 & 3.69109 & -0.50990 \\ 
$10^{9}$ & 2500 & 0.90000 & 3.71067 & -0.38350 \\ 
$10^{9}$ & 2500 & 0.95000 & 3.72878 & -0.25901 \\ 
$10^{9}$ & 2500 & 1.00000 & 3.74464 & -0.13905 \\ 
$10^{9}$ & 2500 & 1.05000 & 3.75891 & -0.02038 \\ 
$10^{9}$ & 2500 & 1.10000 & 3.77392 & 0.12937 \\ 
$10^{9}$ & 2500 & 1.15000 & 3.78460 & 0.24225 \\ 
$10^{9}$ & 2500 & 1.20000 & 3.79403 & 0.35057 \\ 
$10^{9}$ & 2500 & 1.25000 & 3.79990 & 0.41950 \\ 
$10^{9}$ & 2500 & 1.30000 & 3.80672 & 0.51870 \\ 
$10^{9}$ & 2500 & 1.36000 & 3.80966 & 0.62345 \\ 
$10^{9}$ & 2500 & 1.40000 & 3.80714 & 0.68143 \\ 
$10^{9}$ & 2500 & 1.44000 & 3.80200 & 0.73701 \\ 
$10^{9}$ & 2500 & 1.45000 & 3.80077 & 0.75093 \\ 
$10^{9}$ & 2500 & 1.46000 & 3.80014 & 0.76998 \\ 
$10^{9}$ & 2500 & 1.47000 & 3.80387 & 0.80321 \\ 
$10^{9}$ & 2500 & 1.47400 & 3.80837 & 0.82857 \\ 
$10^{9}$ & 2500 & 1.47500 & 3.82174 & 0.89839 \\ 
$10^{9}$ & 2500 & 1.47800 & 3.82165 & 0.90921 \\ 
$10^{9}$ & 2500 & 1.48000 & 3.82125 & 0.91408 \\ 
$10^{9}$ & 2500 & 1.50000 & 3.81726 & 0.96577 \\ 
$10^{9}$ & 2500 & 1.51000 & 3.81292 & 0.99287 \\ 
$10^{9}$ & 2500 & 1.53000 & 3.79977 & 1.03847 \\ 
$10^{9}$ & 2500 & 1.53400 & 3.79405 & 1.04669 \\ 
$10^{9}$ & 2500 & 1.53800 & 3.78823 & 1.05125 \\ 
$10^{9}$ & 2500 & 1.54200 & 3.77927 & 1.04842 \\ 
$10^{9}$ & 2500 & 1.54600 & 3.76432 & 1.02028 \\ 
$10^{9}$ & 2500 & 1.55000 & 3.74863 & 0.96063 \\ 
$10^{9}$ & 2500 & 1.55200 & 3.72995 & 0.87248 \\ 
$10^{9}$ & 2500 & 1.55300 & 3.72165 & 0.85150 \\ 
$10^{9}$ & 2500 & 1.55350 & 3.71679 & 0.85115 \\ 
$10^{9}$ & 2500 & 1.55400 & 3.71266 & 0.86478 \\ 
$10^{9}$ & 2500 & 1.55600 & 3.70693 & 0.91608 \\ 
$10^{9}$ & 2500 & 1.55800 & 3.70279 & 0.98865 \\ 
$10^{9}$ & 2500 & 1.56000 & 3.70049 & 1.04267 \\ 
$10^{9}$ & 2500 & 1.57000 & 3.68788 & 1.34772 \\ 
$10^{9}$ & 2500 & 1.58000 & 3.66395 & 1.80796 \\ 
$10^{9}$ & 2500 & 1.58100 & 3.66014 & 1.87557 \\ 
 & &&&\\
$10^{9}$ & 5000 & 0.75000 & 3.64640 & -0.77180 \\ 
$10^{9}$ & 5000 & 0.80000 & 3.67425 & -0.61878 \\ 
$10^{9}$ & 5000 & 0.85000 & 3.69490 & -0.48554 \\ 
$10^{9}$ & 5000 & 0.90000 & 3.71525 & -0.35117 \\ 
$10^{9}$ & 5000 & 0.95000 & 3.73403 & -0.21566 \\ 
$10^{9}$ & 5000 & 1.00000 & 3.75046 & -0.08181 \\ 
$10^{9}$ & 5000 & 1.05000 & 3.76474 & 0.05411 \\ 
$10^{9}$ & 5000 & 1.10000 & 3.77823 & 0.22569 \\ 
$10^{9}$ & 5000 & 1.15000 & 3.78537 & 0.35412 \\ 
$10^{9}$ & 5000 & 1.17000 & 3.78721 & 0.40756 \\ 
$10^{9}$ & 5000 & 1.20000 & 3.78911 & 0.47904 \\ 
$10^{9}$ & 5000 & 1.22000 & 3.79055 & 0.55209 \\ 
$10^{9}$ & 5000 & 1.23000 & 3.79010 & 0.58454 \\ 
$10^{9}$ & 5000 & 1.25000 & 3.79241 & 0.62423 \\ 
$10^{9}$ & 5000 & 1.27000 & 3.78795 & 0.69228 \\ 
$10^{9}$ & 5000 & 1.27500 & 3.78626 & 0.71037 \\ 
$10^{9}$ & 5000 & 1.28000 & 3.78342 & 0.73052 \\ 
$10^{9}$ & 5000 & 1.28500 & 3.77986 & 0.75002 \\ 
$10^{9}$ & 5000 & 1.29000 & 3.77557 & 0.76661 \\ 
$10^{9}$ & 5000 & 1.29500 & 3.76825 & 0.77843 \\ 
$10^{9}$ & 5000 & 1.30000 & 3.75508 & 0.77224 \\ 
$10^{9}$ & 5000 & 1.30200 & 3.75176 & 0.76711 \\ 
$10^{9}$ & 5000 & 1.30300 & 3.74258 & 0.74291 \\ 
$10^{9}$ & 5000 & 1.30400 & 3.73698 & 0.72577 \\ 
$10^{9}$ & 5000 & 1.30500 & 3.72483 & 0.69252 \\ 
$10^{9}$ & 5000 & 1.30600 & 3.71911 & 0.68500 \\ 
$10^{9}$ & 5000 & 1.30700 & 3.71121 & 0.69561 \\ 
$10^{9}$ & 5000 & 1.30770 & 3.70509 & 0.73789 \\ 
$10^{9}$ & 5000 & 1.30830 & 3.70498 & 0.74000 \\ 
$10^{9}$ & 5000 & 1.30900 & 3.70290 & 0.76802 \\ 
$10^{9}$ & 5000 & 1.31000 & 3.70110 & 0.80051 \\ 
$10^{9}$ & 5000 & 1.31500 & 3.69144 & 1.06699 \\ 
$10^{9}$ & 5000 & 1.32000 & 3.67088 & 1.52995 \\ 
$10^{9}$ & 5000 & 1.32200 & 3.66317 & 1.68432 \\ 
 & &&&\\
$10^{9}$ & 10000 & 0.75000 & 3.65292 & -0.74005 \\ 
$10^{9}$ & 10000 & 0.80000 & 3.67858 & -0.58609 \\ 
$10^{9}$ & 10000 & 0.85000 & 3.70306 & -0.43126 \\ 
$10^{9}$ & 10000 & 0.90000 & 3.72526 & -0.27330 \\ 
$10^{9}$ & 10000 & 0.95000 & 3.74416 & -0.11285 \\ 
$10^{9}$ & 10000 & 0.97000 & 3.75105 & -0.04468 \\ 
$10^{9}$ & 10000 & 1.00000 & 3.75934 & 0.06140 \\ 
$10^{9}$ & 10000 & 1.02000 & 3.76377 & 0.13498 \\ 
$10^{9}$ & 10000 & 1.04000 & 3.76530 & 0.20115 \\ 
$10^{9}$ & 10000 & 1.05000 & 3.76880 & 0.25944 \\ 
$10^{9}$ & 10000 & 1.06000 & 3.77067 & 0.30809 \\ 
$10^{9}$ & 10000 & 1.07000 & 3.76783 & 0.39441 \\ 
$10^{9}$ & 10000 & 1.08000 & 3.76446 & 0.45089 \\ 
$10^{9}$ & 10000 & 1.08300 & 3.76259 & 0.46879 \\ 
$10^{9}$ & 10000 & 1.09000 & 3.75423 & 0.50814 \\ 
$10^{9}$ & 10000 & 1.09200 & 3.75034 & 0.51868 \\ 
$10^{9}$ & 10000 & 1.09600 & 3.73617 & 0.52357 \\ 
$10^{9}$ & 10000 & 1.09900 & 3.71296 & 0.51744 \\ 
$10^{9}$ & 10000 & 1.09930 & 3.71123 & 0.52080 \\ 
$10^{9}$ & 10000 & 1.09970 & 3.70776 & 0.53190 \\ 
$10^{9}$ & 10000 & 1.10000 & 3.70365 & 0.55586 \\ 
$10^{9}$ & 10000 & 1.10050 & 3.70349 & 0.55774 \\ 
$10^{9}$ & 10000 & 1.10100 & 3.69885 & 0.61259 \\ 
$10^{9}$ & 10000 & 1.10300 & 3.69213 & 0.79446 \\ 
$10^{9}$ & 10000 & 1.10500 & 3.68442 & 1.04566 \\ 
$10^{9}$ & 10000 & 1.10550 & 3.68314 & 1.08039 \\ 
$10^{9}$ & 10000 & 1.10600 & 3.67717 & 1.23010 \\ 
 & &&&\\
$10^{10}$ & 25 & 0.80000 & 3.60171 & -0.72426 \\ 
$10^{10}$ & 25 & 0.90000 & 3.62789 & -0.53664 \\ 
$10^{10}$ & 25 & 1.00000 & 3.65060 & -0.37636 \\ 
$10^{10}$ & 25 & 1.10000 & 3.67594 & -0.21872 \\ 
$10^{10}$ & 25 & 1.20000 & 3.70175 & -0.05352 \\ 
$10^{10}$ & 25 & 1.30000 & 3.72801 & 0.12656 \\ 
$10^{10}$ & 25 & 1.40000 & 3.75677 & 0.34375 \\ 
$10^{10}$ & 25 & 1.45000 & 3.77470 & 0.47998 \\ 
$10^{10}$ & 25 & 1.50000 & 3.79134 & 0.60358 \\ 
$10^{10}$ & 25 & 1.55000 & 3.80644 & 0.70678 \\ 
$10^{10}$ & 25 & 1.60000 & 3.82172 & 0.79405 \\ 
$10^{10}$ & 25 & 1.65000 & 3.83954 & 0.87054 \\ 
$10^{10}$ & 25 & 1.70000 & 3.85951 & 0.93735 \\ 
$10^{10}$ & 25 & 1.74000 & 3.87459 & 0.98642 \\ 
$10^{10}$ & 25 & 1.76000 & 3.88176 & 1.00993 \\ 
$10^{10}$ & 25 & 1.80000 & 3.89513 & 1.05477 \\ 
$10^{10}$ & 25 & 1.90000 & 3.92377 & 1.15625 \\ 
$10^{10}$ & 25 & 2.00000 & 3.94699 & 1.24669 \\ 
$10^{10}$ & 25 & 2.10000 & 3.96680 & 1.32970 \\ 
$10^{10}$ & 25 & 2.20000 & 3.98423 & 1.40794 \\ 
$10^{10}$ & 25 & 2.30000 & 3.99997 & 1.48214 \\ 
$10^{10}$ & 25 & 2.40000 & 4.01437 & 1.55285 \\ 
$10^{10}$ & 25 & 2.50000 & 4.02765 & 1.62046 \\ 
$10^{10}$ & 25 & 2.60000 & 4.04003 & 1.68535 \\ 
$10^{10}$ & 25 & 2.70000 & 4.05162 & 1.74769 \\ 
$10^{10}$ & 25 & 2.80000 & 4.06255 & 1.80782 \\ 
$10^{10}$ & 25 & 3.00000 & 4.08279 & 1.92198 \\ 
$10^{10}$ & 25 & 3.10000 & 4.09219 & 1.97628 \\ 
$10^{10}$ & 25 & 3.20000 & 4.10120 & 2.02890 \\ 
& &&&\\
$10^{10}$ & 100 & 0.70000 & 3.59342 & -0.62850 \\ 
$10^{10}$ & 100 & 0.80000 & 3.61265 & -0.50389 \\ 
$10^{10}$ & 100 & 0.90000 & 3.62725 & -0.40982 \\ 
$10^{10}$ & 100 & 1.00000 & 3.64231 & -0.30581 \\ 
$10^{10}$ & 100 & 1.10000 & 3.65872 & -0.20290 \\ 
$10^{10}$ & 100 & 1.20000 & 3.67634 & -0.09159 \\ 
$10^{10}$ & 100 & 1.30000 & 3.69853 & 0.03759 \\ 
$10^{10}$ & 100 & 1.40000 & 3.73490 & 0.24217 \\ 
$10^{10}$ & 100 & 1.45000 & 3.76230 & 0.41585 \\ 
$10^{10}$ & 100 & 1.50000 & 3.78906 & 0.59608 \\ 
$10^{10}$ & 100 & 1.55000 & 3.80725 & 0.71209 \\ 
$10^{10}$ & 100 & 1.60000 & 3.82376 & 0.80198 \\ 
$10^{10}$ & 100 & 1.65000 & 3.84275 & 0.87850 \\ 
$10^{10}$ & 100 & 1.70000 & 3.86286 & 0.94459 \\ 
$10^{10}$ & 100 & 1.76000 & 3.88460 & 1.01625 \\ 
$10^{10}$ & 100 & 1.80000 & 3.89750 & 1.06027 \\ 
$10^{10}$ & 100 & 1.90000 & 3.92475 & 1.16086 \\ 
$10^{10}$ & 100 & 2.00000 & 3.94693 & 1.25173 \\ 
$10^{10}$ & 100 & 2.10000 & 3.96564 & 1.33581 \\ 
$10^{10}$ & 100 & 2.20000 & 3.98181 & 1.41497 \\ 
$10^{10}$ & 100 & 2.30000 & 3.99615 & 1.49024 \\ 
$10^{10}$ & 100 & 2.40000 & 4.00909 & 1.56242 \\ 
$10^{10}$ & 100 & 2.50000 & 4.02099 & 1.63209 \\ 
$10^{10}$ & 100 & 2.60000 & 4.03205 & 1.69954 \\ 
$10^{10}$ & 100 & 2.70000 & 4.04239 & 1.76480 \\ 
$10^{10}$ & 100 & 2.70200 & 4.04259 & 1.76607 \\ 
$10^{10}$ & 100 & 2.80000 & 4.05207 & 1.82829 \\ 
$10^{10}$ & 100 & 3.00000 & 4.06971 & 1.95027 \\ 
$10^{10}$ & 100 & 3.10000 & 4.07771 & 2.00893 \\ 
$10^{10}$ & 100 & 3.20000 & 4.08521 & 2.06651 \\ 
$10^{10}$ & 100 & 3.30000 & 4.09217 & 2.12272 \\ 
$10^{10}$ & 100 & 3.40000 & 4.09866 & 2.17810 \\ 
$10^{10}$ & 100 & 3.50000 & 4.10458 & 2.23242 \\ 
 & &&&\\
$10^{10}$ & 250 & 0.70000 & 3.59349 & -0.62775 \\ 
$10^{10}$ & 250 & 0.80000 & 3.61265 & -0.50389 \\ 
$10^{10}$ & 250 & 0.90000 & 3.62766 & -0.40301 \\ 
$10^{10}$ & 250 & 1.00000 & 3.64230 & -0.30329 \\ 
$10^{10}$ & 250 & 1.10000 & 3.65798 & -0.19969 \\ 
$10^{10}$ & 250 & 1.20000 & 3.67556 & -0.09093 \\ 
$10^{10}$ & 250 & 1.30000 & 3.69671 & 0.03406 \\ 
$10^{10}$ & 250 & 1.40000 & 3.73128 & 0.22720 \\ 
$10^{10}$ & 250 & 1.44000 & 3.75889 & 0.39117 \\ 
$10^{10}$ & 250 & 1.53000 & 3.80126 & 0.67555 \\ 
$10^{10}$ & 250 & 1.55000 & 3.80817 & 0.71798 \\ 
$10^{10}$ & 250 & 1.60000 & 3.82542 & 0.80970 \\ 
$10^{10}$ & 250 & 1.65000 & 3.84492 & 0.88633 \\ 
$10^{10}$ & 250 & 1.70000 & 3.86508 & 0.95286 \\ 
$10^{10}$ & 250 & 1.80000 & 3.89902 & 1.07001 \\ 
$10^{10}$ & 250 & 1.90000 & 3.92480 & 1.17147 \\ 
$10^{10}$ & 250 & 2.00000 & 3.94476 & 1.26280 \\ 
$10^{10}$ & 250 & 2.11000 & 3.96228 & 1.35670 \\ 
$10^{10}$ & 250 & 2.20000 & 3.97481 & 1.43071 \\ 
$10^{10}$ & 250 & 2.30000 & 3.98578 & 1.50889 \\ 
$10^{10}$ & 250 & 2.40000 & 3.99598 & 1.58609 \\ 
$10^{10}$ & 250 & 2.50000 & 4.00488 & 1.66125 \\ 
$10^{10}$ & 250 & 2.60000 & 4.01246 & 1.73526 \\ 
$10^{10}$ & 250 & 2.70000 & 4.01859 & 1.80770 \\ 
$10^{10}$ & 250 & 2.80000 & 4.02310 & 1.87882 \\ 
$10^{10}$ & 250 & 3.00000 & 4.02552 & 2.01842 \\ 
$10^{10}$ & 250 & 3.10000 & 4.02190 & 2.08620 \\ 
$10^{10}$ & 250 & 3.20000 & 4.01430 & 2.15322 \\ 
$10^{10}$ & 250 & 3.22000 & 4.01277 & 2.16722 \\ 
$10^{10}$ & 250 & 3.24000 & 4.01174 & 2.18256 \\ 
$10^{10}$ & 250 & 3.26000 & 4.01327 & 2.20081 \\ 
$10^{10}$ & 250 & 3.28000 & 4.02670 & 2.23334 \\ 
$10^{10}$ & 250 & 3.28500 & 4.04408 & 2.26151 \\ 
$10^{10}$ & 250 & 3.29000 & 4.05639 & 2.29390 \\ 
$10^{10}$ & 250 & 3.29500 & 4.05395 & 2.30856 \\ 
$10^{10}$ & 250 & 3.30000 & 4.05357 & 2.31636 \\ 
$10^{10}$ & 250 & 3.32000 & 4.04968 & 2.34698 \\ 
$10^{10}$ & 250 & 3.33000 & 4.04627 & 2.36266 \\ 
$10^{10}$ & 250 & 3.34000 & 4.04091 & 2.37787 \\ 
$10^{10}$ & 250 & 3.35000 & 4.03370 & 2.39302 \\ 
$10^{10}$ & 250 & 3.36000 & 4.01745 & 2.40485 \\ 
$10^{10}$ & 250 & 3.37000 & 3.98897 & 2.40714 \\ 
$10^{10}$ & 250 & 3.37200 & 3.95120 & 2.38532 \\ 
$10^{10}$ & 250 & 3.37300 & 3.96704 & 2.39705 \\ 
$10^{10}$ & 250 & 3.37350 & 3.92038 & 2.35792 \\ 
$10^{10}$ & 250 & 3.37400 & 3.89185 & 2.32240 \\ 
$10^{10}$ & 250 & 3.37500 & 3.87884 & 2.31104 \\ 
$10^{10}$ & 250 & 3.37600 & 3.87480 & 2.30377 \\ 
$10^{10}$ & 250 & 3.37610 & 3.82323 & 2.23956 \\ 
$10^{10}$ & 250 & 3.37625 & 3.76925 & 2.14724 \\ 
$10^{10}$ & 250 & 3.37630 & 3.75533 & 2.09758 \\ 
$10^{10}$ & 250 & 3.37650 & 3.75340 & 2.08848 \\ 
$10^{10}$ & 250 & 3.37800 & 3.74435 & 2.04126 \\ 
$10^{10}$ & 250 & 3.37890 & 3.72552 & 1.92500 \\ 
$10^{10}$ & 250 & 3.37900 & 3.71631 & 1.89722 \\ 
$10^{10}$ & 250 & 3.37940 & 3.70380 & 1.94489 \\ 
$10^{10}$ & 250 & 3.37970 & 3.69641 & 2.01061 \\ 
$10^{10}$ & 250 & 3.38000 & 3.69639 & 2.01106 \\ 
$10^{10}$ & 250 & 3.38050 & 3.68514 & 2.15606 \\ 
$10^{10}$ & 250 & 3.38100 & 3.66782 & 2.39365 \\ 
 & &&&\\
$10^{10}$ & 500 & 0.70000 & 3.59342 & -0.62850 \\ 
$10^{10}$ & 500 & 0.80000 & 3.61265 & -0.50389 \\ 
$10^{10}$ & 500 & 0.90000 & 3.62767 & -0.40301 \\ 
$10^{10}$ & 500 & 1.00000 & 3.64230 & -0.30251 \\ 
$10^{10}$ & 500 & 1.10000 & 3.65819 & -0.20023 \\ 
$10^{10}$ & 500 & 1.20000 & 3.67554 & -0.09088 \\ 
$10^{10}$ & 500 & 1.30000 & 3.69647 & 0.03344 \\ 
$10^{10}$ & 500 & 1.40000 & 3.73118 & 0.22720 \\ 
$10^{10}$ & 500 & 1.44000 & 3.75827 & 0.38808 \\ 
$10^{10}$ & 500 & 1.53000 & 3.80224 & 0.68268 \\ 
$10^{10}$ & 500 & 1.55000 & 3.80995 & 0.72853 \\ 
$10^{10}$ & 500 & 1.60000 & 3.82760 & 0.82057 \\ 
$10^{10}$ & 500 & 1.65000 & 3.84768 & 0.89791 \\ 
$10^{10}$ & 500 & 1.70000 & 3.86796 & 0.96601 \\ 
$10^{10}$ & 500 & 1.76000 & 3.88850 & 1.03930 \\ 
$10^{10}$ & 500 & 1.80000 & 3.90008 & 1.08457 \\ 
$10^{10}$ & 500 & 1.90000 & 3.92217 & 1.18698 \\ 
$10^{10}$ & 500 & 2.00000 & 3.93750 & 1.28114 \\ 
$10^{10}$ & 500 & 2.10000 & 3.94795 & 1.37030 \\ 
$10^{10}$ & 500 & 2.15000 & 3.95196 & 1.41438 \\ 
$10^{10}$ & 500 & 2.20000 & 3.95509 & 1.45762 \\ 
$10^{10}$ & 500 & 2.30000 & 3.95728 & 1.54300 \\ 
$10^{10}$ & 500 & 2.40000 & 3.95542 & 1.62688 \\ 
$10^{10}$ & 500 & 2.50000 & 3.94693 & 1.70742 \\ 
$10^{10}$ & 500 & 2.55000 & 3.94184 & 1.75045 \\ 
$10^{10}$ & 500 & 2.55600 & 3.94203 & 1.75654 \\ 
$10^{10}$ & 500 & 2.55700 & 3.94212 & 1.75765 \\ 
$10^{10}$ & 500 & 2.55800 & 3.94217 & 1.75880 \\ 
$10^{10}$ & 500 & 2.55900 & 3.94228 & 1.75997 \\ 
$10^{10}$ & 500 & 2.60000 & 3.99163 & 1.88475 \\ 
$10^{10}$ & 500 & 2.62000 & 3.99297 & 1.92062 \\ 
$10^{10}$ & 500 & 2.64000 & 3.99075 & 1.95728 \\ 
$10^{10}$ & 500 & 2.66000 & 3.98326 & 1.99550 \\ 
$10^{10}$ & 500 & 2.68000 & 3.96681 & 2.03021 \\ 
$10^{10}$ & 500 & 2.68200 & 3.95995 & 2.03320 \\ 
$10^{10}$ & 500 & 2.68600 & 3.95410 & 2.03834 \\ 
$10^{10}$ & 500 & 2.68800 & 3.94773 & 2.04005 \\ 
$10^{10}$ & 500 & 2.69200 & 3.92822 & 2.03882 \\ 
$10^{10}$ & 500 & 2.69400 & 3.89373 & 2.01971 \\ 
$10^{10}$ & 500 & 2.69420 & 3.89459 & 2.01915 \\ 
$10^{10}$ & 500 & 2.69440 & 3.89070 & 2.01627 \\ 
$10^{10}$ & 500 & 2.69460 & 3.89580 & 2.02186 \\ 
$10^{10}$ & 500 & 2.69480 & 3.86146 & 1.98866 \\ 
$10^{10}$ & 500 & 2.69500 & 3.86910 & 1.99730 \\ 
$10^{10}$ & 500 & 2.69520 & 3.85675 & 1.98359 \\ 
$10^{10}$ & 500 & 2.69540 & 3.87282 & 2.00053 \\ 
$10^{10}$ & 500 & 2.69560 & 3.87244 & 2.00054 \\ 
$10^{10}$ & 500 & 2.69580 & 3.83047 & 1.95416 \\ 
$10^{10}$ & 500 & 2.69630 & 3.84622 & 1.97231 \\ 
$10^{10}$ & 500 & 2.69660 & 3.78873 & 1.89355 \\ 
$10^{10}$ & 500 & 2.69690 & 3.79320 & 1.90246 \\ 
$10^{10}$ & 500 & 2.69710 & 3.76255 & 1.81118 \\ 
$10^{10}$ & 500 & 2.69730 & 3.75139 & 1.74539 \\ 
$10^{10}$ & 500 & 2.69750 & 3.74936 & 1.72820 \\ 
$10^{10}$ & 500 & 2.69780 & 3.74743 & 1.72034 \\ 
$10^{10}$ & 500 & 2.69795 & 3.73908 & 1.65063 \\ 
$10^{10}$ & 500 & 2.69800 & 3.72122 & 1.54824 \\ 
$10^{10}$ & 500 & 2.69860 & 3.71297 & 1.57353 \\ 
$10^{10}$ & 500 & 2.69880 & 3.69589 & 1.78031 \\ 
$10^{10}$ & 500 & 2.70000 & 3.68929 & 1.88306 \\ 
$10^{10}$ & 500 & 2.70050 & 3.69252 & 1.83360 \\ 
$10^{10}$ & 500 & 2.70070 & 3.68060 & 2.01429 \\ 
 & &&&\\
$10^{10}$ & 1000 & 0.70000 & 3.59343 & -0.62835 \\ 
$10^{10}$ & 1000 & 0.80000 & 3.61265 & -0.50388 \\ 
$10^{10}$ & 1000 & 0.90000 & 3.62771 & -0.40237 \\ 
$10^{10}$ & 1000 & 1.00000 & 3.64227 & -0.30241 \\ 
$10^{10}$ & 1000 & 1.10000 & 3.65805 & -0.19957 \\ 
$10^{10}$ & 1000 & 1.20000 & 3.67568 & -0.09093 \\ 
$10^{10}$ & 1000 & 1.30000 & 3.69727 & 0.03518 \\ 
$10^{10}$ & 1000 & 1.40000 & 3.73102 & 0.22653 \\ 
$10^{10}$ & 1000 & 1.44000 & 3.75889 & 0.39277 \\ 
$10^{10}$ & 1000 & 1.53000 & 3.80564 & 0.70306 \\ 
$10^{10}$ & 1000 & 1.60000 & 3.83200 & 0.84284 \\ 
$10^{10}$ & 1000 & 1.65000 & 3.85341 & 0.92307 \\ 
$10^{10}$ & 1000 & 1.70000 & 3.87285 & 0.99406 \\ 
$10^{10}$ & 1000 & 1.76000 & 3.88904 & 1.06749 \\ 
$10^{10}$ & 1000 & 1.80000 & 3.89652 & 1.11273 \\ 
$10^{10}$ & 1000 & 1.83000 & 3.90033 & 1.14480 \\ 
$10^{10}$ & 1000 & 1.87000 & 3.90370 & 1.18600 \\ 
$10^{10}$ & 1000 & 1.90000 & 3.90493 & 1.21566 \\ 
$10^{10}$ & 1000 & 1.95000 & 3.90401 & 1.26376 \\ 
$10^{10}$ & 1000 & 2.00000 & 3.90141 & 1.31448 \\ 
$10^{10}$ & 1000 & 2.05000 & 3.89407 & 1.36077 \\ 
$10^{10}$ & 1000 & 2.06000 & 3.89241 & 1.37057 \\ 
$10^{10}$ & 1000 & 2.07000 & 3.88761 & 1.37607 \\ 
$10^{10}$ & 1000 & 2.08000 & 3.88835 & 1.38831 \\ 
$10^{10}$ & 1000 & 2.09000 & 3.89412 & 1.41124 \\ 
$10^{10}$ & 1000 & 2.11000 & 3.93148 & 1.48036 \\ 
$10^{10}$ & 1000 & 2.12000 & 3.93149 & 1.49956 \\ 
$10^{10}$ & 1000 & 2.12200 & 3.92995 & 1.54880 \\ 
$10^{10}$ & 1000 & 2.16000 & 3.91064 & 1.62459 \\ 
$10^{10}$ & 1000 & 2.16200 & 3.90691 & 1.62914 \\ 
$10^{10}$ & 1000 & 2.16400 & 3.90554 & 1.63129 \\ 
$10^{10}$ & 1000 & 2.16800 & 3.89927 & 1.63903 \\ 
$10^{10}$ & 1000 & 2.17000 & 3.89056 & 1.64023 \\ 
$10^{10}$ & 1000 & 2.17200 & 3.88933 & 1.64312 \\ 
$10^{10}$ & 1000 & 2.17400 & 3.88375 & 1.64434 \\ 
$10^{10}$ & 1000 & 2.17800 & 3.85624 & 1.63663 \\ 
$10^{10}$ & 1000 & 2.17900 & 3.84939 & 1.63333 \\ 
$10^{10}$ & 1000 & 2.18000 & 3.80882 & 1.59958 \\ 
$10^{10}$ & 1000 & 2.18050 & 3.79955 & 1.58839 \\ 
$10^{10}$ & 1000 & 2.18400 & 3.76397 & 1.48527 \\ 
$10^{10}$ & 1000 & 2.18470 & 3.75255 & 1.42385 \\ 
$10^{10}$ & 1000 & 2.18500 & 3.74623 & 1.37775 \\ 
$10^{10}$ & 1000 & 2.18600 & 3.72835 & 1.26181 \\ 
$10^{10}$ & 1000 & 2.18650 & 3.72407 & 1.24779 \\ 
$10^{10}$ & 1000 & 2.18720 & 3.71856 & 1.25013 \\ 
$10^{10}$ & 1000 & 2.18800 & 3.71212 & 1.28788 \\ 
$10^{10}$ & 1000 & 2.19000 & 3.69462 & 1.56416 \\ 
$10^{10}$ & 1000 & 2.19400 & 3.69228 & 1.60791 \\ 
$10^{10}$ & 1000 & 2.19600 & 3.68734 & 1.69522 \\ 
$10^{10}$ & 1000 & 2.20000 & 3.67855 & 1.84529 \\ 
$10^{10}$ & 1000 & 2.21000 & 3.65633 & 2.19809 \\ 
 & &&&\\
$10^{10}$ & 2500 & 0.70000 & 3.59349 & -0.62780 \\ 
$10^{10}$ & 2500 & 0.80000 & 3.61265 & -0.50388 \\ 
$10^{10}$ & 2500 & 0.90000 & 3.62767 & -0.40301 \\ 
$10^{10}$ & 2500 & 1.00000 & 3.64237 & -0.30378 \\ 
$10^{10}$ & 2500 & 1.10000 & 3.65825 & -0.19978 \\ 
$10^{10}$ & 2500 & 1.35000 & 3.70932 & 0.10914 \\ 
$10^{10}$ & 2500 & 1.40000 & 3.73128 & 0.22828 \\ 
$10^{10}$ & 2500 & 1.44000 & 3.75879 & 0.39584 \\ 
$10^{10}$ & 2500 & 1.53000 & 3.81410 & 0.75816 \\ 
$10^{10}$ & 2500 & 1.57000 & 3.83305 & 0.85851 \\ 
$10^{10}$ & 2500 & 1.60000 & 3.84564 & 0.91821 \\ 
$10^{10}$ & 2500 & 1.63000 & 3.85266 & 0.96669 \\ 
$10^{10}$ & 2500 & 1.66000 & 3.85215 & 1.00609 \\ 
$10^{10}$ & 2500 & 1.70000 & 3.84199 & 1.05047 \\ 
$10^{10}$ & 2500 & 1.72000 & 3.83612 & 1.07915 \\ 
$10^{10}$ & 2500 & 1.72100 & 3.85743 & 1.11808 \\ 
$10^{10}$ & 2500 & 1.72200 & 3.86014 & 1.12466 \\ 
$10^{10}$ & 2500 & 1.72400 & 3.87311 & 1.16259 \\ 
$10^{10}$ & 2500 & 1.72800 & 3.87269 & 1.17369 \\ 
$10^{10}$ & 2500 & 1.73200 & 3.87128 & 1.18566 \\ 
$10^{10}$ & 2500 & 1.73600 & 3.87105 & 1.19157 \\ 
$10^{10}$ & 2500 & 1.74000 & 3.87041 & 1.19929 \\ 
$10^{10}$ & 2500 & 1.75000 & 3.86430 & 1.23096 \\ 
$10^{10}$ & 2500 & 1.76000 & 3.86125 & 1.25416 \\ 
$10^{10}$ & 2500 & 1.76300 & 3.85553 & 1.27278 \\ 
$10^{10}$ & 2500 & 1.76900 & 3.85413 & 1.28327 \\ 
$10^{10}$ & 2500 & 1.77200 & 3.85151 & 1.29184 \\ 
$10^{10}$ & 2500 & 1.77500 & 3.84768 & 1.30132 \\ 
$10^{10}$ & 2500 & 1.78100 & 3.83483 & 1.32116 \\ 
$10^{10}$ & 2500 & 1.78400 & 3.82214 & 1.33031 \\ 
$10^{10}$ & 2500 & 1.78700 & 3.81245 & 1.33483 \\ 
$10^{10}$ & 2500 & 1.78850 & 3.80299 & 1.33387 \\ 
$10^{10}$ & 2500 & 1.79000 & 3.78996 & 1.32675 \\ 
$10^{10}$ & 2500 & 1.79200 & 3.77997 & 1.31361 \\ 
$10^{10}$ & 2500 & 1.79300 & 3.76226 & 1.26238 \\ 
$10^{10}$ & 2500 & 1.79450 & 3.75258 & 1.21969 \\ 
$10^{10}$ & 2500 & 1.79550 & 3.73794 & 1.13340 \\ 
$10^{10}$ & 2500 & 1.79560 & 3.73550 & 1.11770 \\ 
$10^{10}$ & 2500 & 1.79580 & 3.72151 & 1.05788 \\ 
$10^{10}$ & 2500 & 1.79800 & 3.70862 & 1.11460 \\ 
$10^{10}$ & 2500 & 1.80210 & 3.68846 & 1.49512 \\ 
$10^{10}$ & 2500 & 1.80230 & 3.68561 & 1.54939 \\ 
 & &&&\\
$10^{10}$ & 5000 & 0.70000 & 3.59343 & -0.62848 \\
$10^{10}$ & 5000 & 0.80000 & 3.61266 & -0.50388 \\
$10^{10}$ & 5000 & 0.90000 & 3.62768 & -0.40294 \\
$10^{10}$ & 5000 & 1.00000 & 3.64232 & -0.30238 \\
$10^{10}$ & 5000 & 1.10000 & 3.65825 & -0.19996 \\
$10^{10}$ & 5000 & 1.35000 & 3.71131 & 0.11569 \\
$10^{10}$ & 5000 & 1.40000 & 3.73269 & 0.23555 \\
$10^{10}$ & 5000 & 1.42000 & 3.74773 & 0.32335 \\
$10^{10}$ & 5000 & 1.44000 & 3.75940 & 0.40507 \\
$10^{10}$ & 5000 & 1.52500 & 3.82673 & 0.86081 \\ 
$10^{10}$ & 5000 & 1.52900 & 3.82723 & 0.87224 \\ 
$10^{10}$ & 5000 & 1.53000 & 3.82739 & 0.87449 \\ 
$10^{10}$ & 5000 & 1.53200 & 3.82716 & 0.87939 \\ 
$10^{10}$ & 5000 & 1.55000 & 3.82059 & 0.91006 \\ 
$10^{10}$ & 5000 & 1.55400 & 3.81926 & 0.91844 \\ 
$10^{10}$ & 5000 & 1.55500 & 3.81942 & 0.92292 \\ 
$10^{10}$ & 5000 & 1.55600 & 3.82012 & 0.92808 \\ 
$10^{10}$ & 5000 & 1.55700 & 3.82209 & 0.93831 \\ 
$10^{10}$ & 5000 & 1.55800 & 3.82537 & 0.95100 \\ 
$10^{10}$ & 5000 & 1.56000 & 3.83293 & 0.98387 \\ 
$10^{10}$ & 5000 & 1.57000 & 3.82934 & 1.02482 \\ 
$10^{10}$ & 5000 & 1.57400 & 3.82722 & 1.04105 \\ 
$10^{10}$ & 5000 & 1.57800 & 3.82450 & 1.06319 \\ 
$10^{10}$ & 5000 & 1.58200 & 3.82054 & 1.08576 \\ 
$10^{10}$ & 5000 & 1.58600 & 3.81474 & 1.10958 \\ 
$10^{10}$ & 5000 & 1.59000 & 3.80618 & 1.13377 \\ 
$10^{10}$ & 5000 & 1.59200 & 3.80030 & 1.14496 \\ 
$10^{10}$ & 5000 & 1.59400 & 3.79237 & 1.15433 \\ 
$10^{10}$ & 5000 & 1.59600 & 3.78150 & 1.15645 \\ 
$10^{10}$ & 5000 & 1.59700 & 3.77129 & 1.14679 \\ 
$10^{10}$ & 5000 & 1.59800 & 3.76298 & 1.12934 \\ 
$10^{10}$ & 5000 & 1.59850 & 3.75192 & 1.09169 \\ 
$10^{10}$ & 5000 & 1.59900 & 3.74191 & 1.04429 \\ 
$10^{10}$ & 5000 & 1.59930 & 3.73466 & 1.00691 \\ 
$10^{10}$ & 5000 & 1.59970 & 3.73375 & 1.00286 \\ 
$10^{10}$ & 5000 & 1.60000 & 3.72418 & 0.96280 \\ 
$10^{10}$ & 5000 & 1.60020 & 3.71859 & 0.95378 \\ 
$10^{10}$ & 5000 & 1.60040 & 3.70956 & 0.98761 \\ 
$10^{10}$ & 5000 & 1.60060 & 3.70970 & 0.98616 \\ 
$10^{10}$ & 5000 & 1.60070 & 3.70557 & 1.03330 \\ 
$10^{10}$ & 5000 & 1.60100 & 3.70588 & 1.02912 \\ 
$10^{10}$ & 5000 & 1.60200 & 3.69679 & 1.20695 \\ 
$10^{10}$ & 5000 & 1.60300 & 3.68802 & 1.39689 \\ 
$10^{10}$ & 5000 & 1.60400 & 3.66942 & 1.74128 \\ 
$10^{10}$ & 5000 & 1.60500 & 3.65370 & 2.00792 \\ 
$10^{10}$ & 5000 & 1.60600 & 3.64008 & 2.21617 \\
& &&&\\
$10^{10}$ & 10000 & 0.70000 & 3.59344 & -0.62847 \\ 
$10^{10}$ & 10000 & 0.80000 & 3.61267 & -0.50388 \\ 
$10^{10}$ & 10000 & 0.90000 & 3.62769 & -0.40299 \\ 
$10^{10}$ & 10000 & 1.30000 & 3.69722 & 0.03529 \\ 
$10^{10}$ & 10000 & 1.40000 & 3.73533 & 0.24953 \\ 
$10^{10}$ & 10000 & 1.41000 & 3.74352 & 0.29694 \\ 
$10^{10}$ & 10000 & 1.46500 & 3.80456 & 0.70287 \\ 
$10^{10}$ & 10000 & 1.48000 & 3.81716 & 0.82561 \\ 
$10^{10}$ & 10000 & 1.48200 & 3.81629 & 0.84311 \\
$10^{10}$ & 10398\footnote[1]{In the case of $\rho_{\chi}=10^{10}\;$GeV cm$^{-3}$, the post-MS segment of the 10 Gyr isochrone is a conservative estimation (a lower limit on luminosity) of the true isochrone. It corresponds to the evolutionary track of a star of 1.482 M$_{\odot}$.} & 1.48200 & 3.81542 & 0.86725 \\ 
$10^{10}$ & 11050\footnotemark[1] & 1.48200 & 3.81191 & 0.92592 \\ 
$10^{10}$ & 11480\footnotemark[1] & 1.48200 & 3.80361 & 0.98288 \\ 
$10^{10}$ & 11531\footnotemark[1] & 1.48200 & 3.80181 & 0.99086 \\ 
$10^{10}$ & 11764\footnotemark[1] & 1.48200 & 3.78869 & 1.02901 \\ 
$10^{10}$ & 11859\footnotemark[1] & 1.48200 & 3.77838 & 1.04011 \\ 
$10^{10}$ & 11886\footnotemark[1] & 1.48200 & 3.77429 & 1.04049 \\ 
$10^{10}$ & 11911\footnotemark[1] & 1.48200 & 3.76976 & 1.03835 \\ 
$10^{10}$ & 11953\footnotemark[1] & 1.48200 & 3.75900 & 1.02202 \\ 
$10^{10}$ & 11989\footnotemark[1] & 1.48200 & 3.74529 & 0.97853 \\ 
$10^{10}$ & 12011\footnotemark[1] & 1.48200 & 3.73269 & 0.92437 \\ 
$10^{10}$ & 12023\footnotemark[1] & 1.48200 & 3.72535 & 0.89746 \\ 
$10^{10}$ & 12030\footnotemark[1] & 1.48200 & 3.72109 & 0.88802 \\ 
$10^{10}$ & 12043\footnotemark[1] & 1.48200 & 3.71376 & 0.89170 \\ 
$10^{10}$ & 12055\footnotemark[1] & 1.48200 & 3.70812 & 0.92433 \\ 
$10^{10}$ & 12063\footnotemark[1] & 1.48200 & 3.70488 & 0.96159 \\ 
$10^{10}$ & 12073\footnotemark[1] & 1.48200 & 3.70197 & 1.00946 \\ 
$10^{10}$ & 12091\footnotemark[1] & 1.48200 & 3.69698 & 1.11706 \\ 
$10^{10}$ & 12101\footnotemark[1] & 1.48200 & 3.69396 & 1.18796 \\ 
$10^{10}$ & 12115\footnotemark[1] & 1.48200 & 3.68955 & 1.28798 \\ 
$10^{10}$ & 12125\footnotemark[1] & 1.48200 & 3.68606 & 1.36247 \\ 
$10^{10}$ & 12135\footnotemark[1] & 1.48200 & 3.68178 & 1.44963 \\ 
$10^{10}$ & 12146\footnotemark[1] & 1.48200 & 3.67543 & 1.56985 \\ 
$10^{10}$ & 12157\footnotemark[1] & 1.48200 & 3.66829 & 1.69748 \\ 
$10^{10}$ & 12161\footnotemark[1] & 1.48200 & 3.66535 & 1.74837 \\ 
$10^{10}$ & 12173\footnotemark[1] & 1.48200 & 3.66014 & 1.84078 \\ 
$10^{10}$ & 12184\footnotemark[1] & 1.48200 & 3.65567 & 1.91832 \\ 
$10^{10}$ & 12189\footnotemark[1] & 1.48200 & 3.64898 & 2.02439 \\ 
\label{tab-alliso}
\end{longtable}

\begin{thebibliography}{50}
\expandafter\ifx\csname natexlab\endcsname\relax\def\natexlab#1{#1}\fi

\bibitem[{Ahmed} {et~al.}(2010)]{art-CDMS2010Sci} 
{Ahmed}, Z. {et~al.} (CDMS II Collaboration) 2010, Science, 327, 1619-, arXiv:0912.3592

\bibitem[{{Bartko} {et~al.}(2010){Bartko}, {Martins}, {Trippe}, {Fritz},
  {Genzel}, {Ott}, {Eisenhauer}, {Gillessen}, {Paumard}, {Alexander},
  {Dodds-Eden}, {Gerhard}, {Levin}, {Mascetti}, {Nayakshin}, {Perets},
  {Perrin}, {Pfuhl}, {Reid}, {Rouan}, {Zilka}, \&
  {Sternberg}}]{art-Bartkoetal2010ApJ}
{Bartko}, H. {et~al.} 2010, \apj, 708, 834, arXiv:0908.2177

\bibitem[{{Behnke} {et~al.}(2011){Behnke}, {Behnke}, {Brice}, {Broemmelsiek},
  {Collar}, {Cooper}, {Crisler}, {Dahl}, {Fustin}, {Hall}, {Hinnefeld}, {Hu},
  {Levine}, {Ramberg}, {Shepherd}, {Sonnenschein}, \&
  {Szydagis}}]{art-COUPP2011PhRvL}
{Behnke}, E. {et~al.} (COUPP Collaboration) 2011, Physical Review Letters, 106,
  021303, arXiv:1008.3518

\bibitem[{{Bernal} \& {Palomares-Ruiz}(2010)}]{art-BernalPalomares2010}
{Bernal}, N. \& {Palomares-Ruiz}, S. 2010, arXiv:1006.0477

\bibitem[{{Bertone}(2010)}]{art-Bertone2010Nature}
{Bertone}, G. 2010, \nat, 468, 389, arXiv:1011.3532

\bibitem[{Bertone {et~al.}(2010)Bertone, Cerdeno, Fornasa, de~Austri, \&
  Trotta}]{art-Bertone2010PhRevD}
Bertone, G., Cerdeno, D.~G., Fornasa, M., de~Austri, R.~R., \& Trotta, R. 2010,
  Phys. Rev., D82, 055008, arXiv:1005.4280

\bibitem[{{Bertone} \& {Fairbairn}(2008)}]{art-BertoneFairbairn2008}
{Bertone}, G. \& {Fairbairn}, M. 2008, Phys. Rev. D, 77, 043515, arXiv:0709.1485

\bibitem[{{Bertone} \& {Merritt}(2005)}]{art-BertoneMerritt2005}
{Bertone}, G. \& {Merritt}, D. 2005, Modern Physics Letters A, 20, 1021, arXiv:astro-ph/0504422

\bibitem[{{Buchholz} {et~al.}(2009){Buchholz}, {Sch{\"o}del}, \&
  {Eckart}}]{art-Buchholzetal2009A&A}
{Buchholz}, R.~M., {Sch{\"o}del}, R., \& {Eckart}, A. 2009, \aap, 499, 483, arXiv:0903.2135

\bibitem[{{Carraro} {et~al.}(2002){Carraro}, {Girardi}, \&
  {Marigo}}]{art-Carraroetal2002MNRAS}
{Carraro}, G., {Girardi}, L., \& {Marigo}, P. 2002, \mnras, 332, 705, arXiv:astro-ph/0202018 

\bibitem[{Casanellas \& Lopes(2009)}]{art-CasanellasLopes2009ApJ}
Casanellas, J. \& Lopes, I. 2009, ApJ, 705, 135, arXiv:0909.1971

\bibitem[{Casanellas \& Lopes(2011)}]{art-CasanellasLopes2010MNRAS}
Casanellas, J. \& Lopes, I. 2011, MNRAS, 410, 535, arXiv:1008.0646

\bibitem[{{Cerde{\~n}o} \& {Green}(2010)}]{chap-CerdenoGreen2010}
{Cerde{\~n}o}, D.~G. \& {Green}, A.~M. {Direct detection of WIMPs}, ed.
  {Bertone, G.} (Cambridge University Press), 347--+, arXiv:1002.1912

\bibitem[{Cumberbatch {et~al.}(2010)Cumberbatch, Guzik, Silk, Watson, \&
  West}]{art-Cumberbatchetal2010PhRevD}
Cumberbatch, D.~T., Guzik, J., Silk, J., Watson, L.~S., \& West, S.~M. 2010,
  Phys. Rev., D82, 103503, arXiv:1005.5102

\bibitem[{de Blok (2010)}]{art-deBlok2010}
de Blok, W. J. G. 2010, Advances in Astronomy, vol. 2010, 789293, doi:10.1155/2010/789293, arXiv:0910.3538

\bibitem[{{de Lavallaz} \& {Fairbairn}(2010)}]{art-LavallazFairbairn2010PhRvD}
{de Lavallaz}, A. \& {Fairbairn}, M. 2010, Phys. Rev. D, 81, 123521, arXiv:1004.0629

\bibitem[{{Do} {et~al.}(2009){Do}, {Ghez}, {Morris}, {Lu}, {Matthews}, {Yelda},
  \& {Larkin}}]{art-Doetal2009ApJ}
{Do}, T., {Ghez}, A.~M., {Morris}, M.~R., {Lu}, J.~R., {Matthews}, K., {Yelda},
  S., \& {Larkin}, J. 2009, \apj, 703, 1323, arXiv:0908.0311

\bibitem[{{Fairbairn} {et~al.}(2008){Fairbairn}, {Scott}, \&
  {Edsj{\"{o}}}}]{art-Fairbairnetal2008}
{Fairbairn}, M., {Scott}, P., \& {Edsj{\"{o}}}, J. 2008, Phys. Rev. D, 77, 047301, arXiv:0710.3396

\bibitem[{{Girardi} {et~al.}(2000){Girardi}, {Bressan}, {Bertelli}, \&
  {Chiosi}}]{art-Girardietal2000A&AS}
{Girardi}, L., {Bressan}, A., {Bertelli}, G., \& {Chiosi}, C. 2000, \aaps, 141,
  371, arXiv:astro-ph/9910164

\bibitem[{{Gondolo} {et~al.}(2004){Gondolo}, {Edsj{\"{o}}}, {Ullio},
  {Bergstr{\"{o}}m}, {Schelke}, \& {Baltz}}]{art-GondoloEdsjoDarkSusy2004}
{Gondolo}, P., {Edsj{\"{o}}}, J., {Ullio}, P., {Bergstr{\"{o}}m}, L.,
  {Schelke}, M., \& {Baltz}, E.~A. 2004, JCAP, 7, 008, arXiv:astro-ph/0406204

\bibitem[{{Gondolo} {et~al.}(2010){Gondolo}, {Huh}, {Do Kim}, \&
  {Scopel}}]{art-GondoloHuhKimScopel2010JCAP}
{Gondolo}, P., {Huh}, J., {Do Kim}, H., \& {Scopel}, S. 2010, JCAP, 7, 26, arXiv:1004.1258

\bibitem[{{Gondolo} \& {Silk}(1999)}]{art-GondoloSilk1999}
{Gondolo}, P. \& {Silk}, J. 1999, Phys. Rev. Lett., 83, 1719, arXiv:astro-ph/9906391

\bibitem[{{Gould}(1987)}]{art-Gould1987}
{Gould}, A. 1987, ApJ, 321, 571

\bibitem[{{Iocco}(2008)}]{art-Iocco2008}
{Iocco}, F. 2008, ApJ, 677, L1, arXiv:0802.0941

\bibitem[{{Iocco} {et~al.}(2008){Iocco}, {Bressan}, {Ripamonti}, {Schneider},
  {Ferrara}, \& {Marigo}}]{art-Ioccoetal2008}
{Iocco}, F., {Bressan}, A., {Ripamonti}, E., {Schneider}, R., {Ferrara}, A., \&
  {Marigo}, P. 2008, MNRAS, 390, 1655, arXiv:0805.4016

\bibitem[{{Kouvaris} \& {Tinyakov}(2010)}]{art-KouvarisTinyakov2010}
{Kouvaris}, C. \& {Tinyakov}, P. 2011, Phys. Rev. D, 83, 083512, arXiv:1012.2039

\bibitem[{Lopes, Casanellas \& Eug\'enio (2011)}]{art-LopCasEug2011PhRevD}
Lopes, I., Casanellas, J. \& Eug\'enio, D. 2011, Phys. Rev. D, 83, 063521, arXiv:1102.2907

\bibitem[{Lopes \& Silk(2010a)}]{art-LopesSilk2010Science}
Lopes, I. \& Silk, J. 2010a, Science, 330, 462

\bibitem[{{Lopes} \& {Silk}(2010b)}]{art-LopesSilk2010ApJL}
{Lopes}, I. \& {Silk}, J. 2010b, \apjl, 722, L95, arXiv:1009.5122

\bibitem[Lu et~al.(2009)]{art-LuGhezetal2009}
Lu, J. R.   m*-+al. 2009, ApJ, 690, 1463, arXiv:0808.3818

\bibitem[{{Morel}(1997)}]{art-Morel1997}
{Morel}, P. 1997, A\&As, 124, 597, 

\bibitem[{{Moskalenko} \& {Wai}(2007)}]{art-MoskalenkoWai2007}
{Moskalenko}, I.~V. \& {Wai}, L.~L. 2007, ApJ, 659, L29, arXiv:astro-ph/0702654

\bibitem[{Pato {et~al.}(2010)Pato, Baudis, Bertone, Ruiz~de Austri, Strigari,
  \& Trotta}]{art-Patoetal2010arXiv}
Pato, M., Baudis, L., Bertone, G., Ruiz~de Austri, R., Strigari, L.~E., \&
  Trotta, R. 2011, Phys. Rev. D, 83, 083505, arXiv:1012.3458

\bibitem[{{Ripamonti} {et~al.}(2010){Ripamonti}, {Iocco}, {Ferrara},
  {Schneider}, {Bressan}, \& {Marigo}}]{art-RipamontiIocooetal2010MNRAS}
{Ripamonti}, E., {Iocco}, F., {Ferrara}, A., {Schneider}, R., {Bressan}, A., \&
  {Marigo}, P. 2010, MNRAS, 406, 2605, arXiv:1003.0676

\bibitem[{Salati \& Silk (1989)}]{art-SalatiSilk1989}
{Salati}, P. \& {Silk}, J. 1989, \apj, 338, 24

\bibitem[{{Savage} {et~al.}(2009){Savage}, {Gelmini}, {Gondolo}, \&
  {Freese}}]{art-Savageetal2009JCAP}
{Savage}, C., {Gelmini}, G., {Gondolo}, P., \& {Freese}, K. 2009, JCAP, 4, 010, arXiv:0808.3607

\bibitem[{{Scott} {et~al.}(2010){Scott}, {Conrad}, {Edsj{\"o}},
  {Bergstr{\"o}m}, {Farnier}, \& {Akrami}}]{art-Scottetal2010JCAP}
{Scott}, P., {Conrad}, J., {Edsj{\"o}}, J., {Bergstr{\"o}m}, L., {Farnier}, C.,
  \& {Akrami}, Y. 2010, JCAP, 1, 31, arXiv:0909.3300 

\bibitem[{{Scott} {et~al.}(2009){Scott}, {Fairbairn}, \&
  {Edsj{\"{o}}}}]{art-Scottetal2009MNRAS}
{Scott}, P., {Fairbairn}, M., \& {Edsj{\"{o}}}, J. 2009, MNRAS, 394, 82, arXiv:0809.1871

\bibitem[{Sivertsson \& Gondolo(2010)}]{art-SivertssonGondolo2010}
Sivertsson, S. \& Gondolo, P. 2011, ApJ, 729, 51, arXiv:1006.0025

\bibitem[{{Spolyar} {et~al.}(2008){Spolyar}, {Freese}, \&
  {Gondolo}}]{let-Spolyaretal2008}
{Spolyar}, D., {Freese}, K., \& {Gondolo}, P. 2008, Phys. Rev. Lett., 100, 051101, arXiv:0705.0521

\bibitem[{Taoso {et~al.}(2008)Taoso, Bertone, Meynet, \&
  Ekstr\"{o}m}]{art-Taosoetal2008PhRevD}
Taoso, M., Bertone, G., Meynet, G., \& Ekstr\"{o}m, S. 2008, Phys. Rev. D, 78, 123510, arXiv:0806.2681

\bibitem[{{Taoso} {et~al.}(2010){Taoso}, {Iocco}, {Meynet}, {Bertone}, \&
  {Eggenberger}}]{art-Taosoetal2010PhRvD}
{Taoso}, M., {Iocco}, F., {Meynet}, G., {Bertone}, G., \& {Eggenberger}, P.
  2010, \prd, 82, 083509, arXiv:1005.5711

\bibitem[{{Trotta} {et~al.}(2009){Trotta}, {Ruiz de Austri}, \& {P{\'e}rez de
  los Heros}}]{art-Trottaetal2009JCAP}
{Trotta}, R., {Ruiz de Austri}, R., \& {P{\'e}rez de los Heros}, C. 2009, JCAP,
  8, 34, arXiv:0906.0366

\bibitem[{{Turck-Chieze} \& {Lopes}(1993)}]{art-TurckLopes1993ApJ}
{Turck-Chieze}, S. \& {Lopes}, I. 1993, \apj, 408, 347, 

\bibitem[{{Turck-Chi{\`{e}}ze} {et~al.}(2010){Turck-Chi{\`{e}}ze}, {Palacios},
  {Marques}, \& {Nghiem}}]{art-Turck-Chiezeetal2010ApJ}
{Turck-Chi{\`{e}}ze}, S., {Palacios}, A., {Marques}, J.~P., \& {Nghiem}, P.
  A.~P. 2010, ApJ, 715, 1539, arXiv:1004.1657

\bibitem[{{Twarog} {et~al.}(2011){Twarog}, {Carraro}, \&
  {Anthony-Twarog}}]{art-Twarogetal2011ApJL}
{Twarog}, B.~A., {Carraro}, G., \& {Anthony-Twarog}, B.~J. 2011, \apjl, 727,
  L7+, arXiv:1011.5138

\bibitem[{{Yoon} {et~al.}(2008){Yoon}, {Iocco}, \&
  {Akiyama}}]{art-YoonIoccoAkiyama2008}
{Yoon}, S.~C., {Iocco}, F., \& {Akiyama}, S. 2008, ApJ, 688, L1, arXiv:0806.2662

\bibitem[{{Yuan} {et~al.}(2011){Yuan}, {Yue}, {Zhang}, \& {Chen}}]{art-Yuanetal2011}
{Yuan}, Q., {Yue}, B., {Zhang}, B., \& {Chen}, X. 2011, JCAP, 4, 020, arXiv:1104.1233

\bibitem[{Zackrisson} {et~al.}(2010)]{art-ZackScottIoccoetal2010ApJ}
{Zackrisson}, E. {et~al.} 2010, ApJ, 717, 257, arXiv:1002.3368

\bibitem[{{Zhao} \& {Bailyn}(2005)}]{art-Zhao2005AJ}
{Zhao}, B. \& {Bailyn}, C.~D. 2005, \aj, 129, 1934, 

\end{thebibliography}
\end{document}